\documentclass[useAMS,usenatbib]{mn2e}

\usepackage[pass]{geometry}
\usepackage{graphicx}

\title[Outer Morphology of S$^{4}$G Galaxies]{{\it Spitzer}/IRAC Near-Infrared Features in the Outer Parts of 
S$^{4}$G Galaxies}
\author[S. Laine et al.]{Seppo Laine,$^{1}$\thanks{Email: seppo@ipac.caltech.edu}
Johan H. Knapen,$^{2,3}$ Juan--Carlos Mu\~{n}oz--Mateos,$^{4.5}$
Taehyun \newauthor Kim,$^{4,5,6,7}$ S\'{e}bastien Comer\'{o}n,$^{8,9}$ Marie Martig,$^{10}$
Benne W. Holwerda,$^{11}$ \newauthor E. Athanassoula,$^{12}$ Albert Bosma,$^{12}$
Peter H. Johansson,$^{13}$ \newauthor Santiago Erroz--Ferrer,$^{2,3}$ Dimitri A. Gadotti,$^{5}$
Armando Gil de Paz,$^{14}$ \newauthor Joannah Hinz,$^{15}$ Jarkko Laine,$^{8,9}$
Eija Laurikainen,$^{8,9}$ \newauthor Kar\'{i}n Men\'{e}ndez--Delmestre,$^{16}$ Trisha Mizusawa,$^{4,17}$
Michael W. Regan,$^{18}$ \newauthor Heikki Salo,$^{8}$ Kartik Sheth,$^{4,1,19}$ Mark Seibert,$^{7}$
Ronald J. Buta,$^{20}$ \newauthor Mauricio Cisternas,$^{2,3}$
Bruce G. Elmegreen,$^{21}$ Debra M. Elmegreen,$^{22}$ \newauthor Luis C. Ho,$^{23,7}$ Barry F. Madore$^{7}$
and Dennis Zaritsky$^{24}$\\
$^{1}${\it Spitzer} Science Center - Caltech, MS 314-6, Pasadena, CA 91125, USA\\
$^{2}$Instituto de Astrof\'{i}sica de Canarias, E-38205 La Laguna, Tenerife, Spain\\
$^{3}$Departamento de Astrof\'{i}sica, Universidad de La Laguna, 38206 La Laguna, Spain\\
$^{4}$National Radio Astronomy Observatory/NAASC, Charlottesville, 520 Edgemont Road, VA 22903, USA\\
$^{5}$European Southern Observatory, Alonso de Cordova 3107, Vitacura, Casilla 19001, Santiago, Chile\\
$^{6}$Astronomy Program, Department of Physics and Astronomy, Seoul National University, Seoul 151-742, Korea\\
$^{7}$The Observatories of the Carnegie Institution of Washington, 813 Santa Barbara Street, Pasadena, CA 91101, USA\\
$^{8}$Division of Astronomy, Department of Physics, University of Oulu, P.O. Box 3000, 90014 Oulu, Finland\\
$^{9}$Finnish Centre of Astronomy with ESO (FINCA), University of Turku, V\"{a}is\"{a}l\"{a}ntie 20, FIN-21500 Piikki\"{o}\\
$^{10}$Max-Planck Institut f\"{u}r Astronomie, K\"{o}nigstuhl 17 D-69117 Heidelberg, Germany\\
$^{11}$Leiden Observatory, Leiden University, P.O. Box 9513, 2300 RA Leiden, The Netherlands\\
$^{12}$Aix Marseille Universit\'{e}, CNRS, LAM (Laboratoire d'Astrophysique de Marseille), UMR 7326, 13388 Marseille 13, France \\
$^{13}$Department of Physics, University of Helsinki, Gustaf H\"{a}llstr\"{o}min katu 2a, 00014 Helsinki, Finland\\
$^{14}$Departamento de Astrof\'{i}sica y CC. de la Atm\'{o}sfera, Universidad Complutense de Madrid, Avda. de la Complutense s/n,\\ Madrid E-28040, Spain\\
$^{15}$MMTO, University of Arizona, 933 N. Cherry Avenue, Tucson, AZ 85721, USA\\
$^{16}$Universidade Federal do Rio de Janeiro, Observat{\'o}rio do Valongo, Ladeira do Pedro Ant{\^{o}}nio, 43, CEP 20080-090, 
\\ Rio de Janeiro, Brazil\\
$^{17}$Florida Institute of Technology, Melbourne, FL 32901\\
$^{18}$Space Telescope Science Institute, 3700 San Martin Drive, Baltimore, MD 21218, USA\\
$^{19}$California Institute of Technology, 1200 East California Boulevard, Pasadena, CA 91125\\
$^{20}$Department of Physics and Astronomy, University of Alabama, Box 870324, Tuscaloosa, AL 35487, USA\\
$^{21}$IBM Research Division, T. J. Watson Research Center, 1101 Kitchawan Road, Yorktown Heights, NY 10598, USA\\
$^{22}$Department of Physics and Astronomy, Vassar College, Poughkeepsie, NY 12604, USA\\
$^{23}$Kavli Institute for Astronomy and Astrophysics, Peking University, Beijing 100871, China\\
$^{24}$Steward Observatory, University of Arizona, 933 North Cherry Avenue, Tucson, AZ 85721, USA}

\begin{document}

\input psfig.sty

\date{To be Published in MNRAS}

\pagerange{\pageref{firstpage}--\pageref{lastpage}} \pubyear{2014}

\maketitle

\label{firstpage}

\clearpage

\begin{abstract} {We present a catalogue and images of visually detected features,
such as asymmetries, extensions, warps, shells, tidal tails, polar rings, and obvious 
signs of mergers or interactions, in the faint outer regions (at and outside of
$R_{\rm 25}$) of nearby galaxies. This catalogue can be used in future quantitative
studies that examine galaxy evolution due to internal and external factors. We are 
able to reliably detect outer region features down to a 
brightness level of 0.03 MJy/sr per pixel at 3.6 $\mu$m in the {\it Spitzer} Survey 
of Stellar Structure in Galaxies (S$^{4}$G). 
We also tabulate companion galaxies. We find asymmetries in the outer
isophotes in 22$\pm1$ per cent of the sample. The asymmetry fraction does not
correlate with galaxy classification as an interacting galaxy or merger
remnant, or with the presence of companions. We also compare the detected features to
similar features in galaxies taken from cosmological zoom re-simulations. 
The simulated images have a higher fraction (33 per cent) of outer 
disc asymmetries, which may be due to selection effects and an uncertain star formation 
threshold in the models. The asymmetries may have either an internal (e.g.,
lopsidedness due to dark halo asymmetry) or external origin.}
\end{abstract}


\section{Introduction}

\label{intro}

Performing studies of the internal or external factors that cause 
galaxies to evolve is predicated on the availability of statistically significant 
numbers of target galaxies that exhibit resolvable, implicative signs of these 
processes. One way to do this is by observing a large sample of nearby galaxies 
and searching for faint features that exist at or outside their outer 
`edges' (at or outside the 25~mag~arcsec$^{-2}$ $B$-band isophotes; `$R_{25}$'). 
Such features may be a sign of past interactions and mergers 
that the targeted galaxy has undergone in its recent or even extended (billions 
of years) past \citep[e.g.,][]{arp66,toomre72,voron59,voron77,hibbard99,hibbard01}. Gas accretion from 
the intragalactic medium, possibly from filaments, may be the cause for faint outer 
features such as warps and polar rings, in addition to asymmetry 
\citep*[e.g.,][]{ostriker89,bournaud03,maccio06,brook08,jog09}.
Internal causes for asymmetry include lopsidedness due to dark halo asymmetry
\citep[e.g,][]{jog09,zaritsky13}. Therefore, statistics of the frequency of existence 
of these features around nearby galaxies will help us to assess the importance 
of the afore-mentioned processes on galaxy evolution.

The main approach to detecting faint features in the outer regions of galaxies 
is through visual classification \citep[e.g.,][and references therein]{sandage05}. 
One of the most well-known catalogues of unusual features in and around galaxies is 
`Arp's Atlas of Peculiar Galaxies' \citep{arp66}. Another fundamentally 
important visual classification of interacting and merging galaxies was made 
by \citet{toomre72}. Other more recent attempts to visually classify galaxy
morphology include `The de Vaucouleurs Atlas of Galaxies' \citep*{buta07} 
and `Galaxy  Morphology' \citep{buta13}. The quantitative approach to detecting 
unusual galaxy features based on, e.g., asymmetry, concentration, clumpiness, and 
the Gini inequality parameter \citep*[e.g.,][]{abraham00,bershady00,abraham03,conselice03,lotz04,scarlata07,
munoz09,holwerda11,huertas13,holwerda14} works better in regions of high signal-to-noise (S/N),
namely, in the inner regions of galaxies. Thus, these two approaches are often 
complementary, as the quantitative method will miss faint features at or outside 
the outer edges of galaxies, where the eye can pick up features 
\citep[e.g.,][]{adams12,hoyos12} that can form the basis for future quantitative 
studies after much deeper, high S/N images are available. Indeed,
when detecting features in the outermost regions of galaxies (or outside their 
continuous luminous bodies), such as outer disc asymmetries, warps,
tidal features, etc., it can be argued that the human eye is still often the 
most effective tool for picking up faint patterns (although attention
needs to be paid to erroneous identifications, such as faint residual images). 
False positive detections can be
reduced to some degree by using more than one person to detect the features of
any given galaxy. The effort to avoid false positive detections, although not in
the context of faint outer features, has been taken to its extreme in the Galaxy
Zoo project \citep[www.galaxyzoo.org;][]{lintott08,lintott11}, which allows
anyone to go online and categorize a shown galaxy with references to a few
illustrated morphological choices. An automated detection
and classification of galaxy features with the help of neural networks has also 
been attempted \citep[e.g.,][]{storrie92,lahav95,goderya02,ball04,fukugita07,ball08,shamir09,cheng11}, 
but so far it has worked better in assigning galaxies into broad morphological
classes based on inner large-scale features, rather than in detecting weak 
patterns outside the main bodies of galaxies. Because any remaining image artefacts
are more prominent outside the main bodies of galaxies, any automatic feature
detections there would likely have to be checked by eye, further reducing the 
usefulness of automatic detection methods outside $R_{25}$.

The visual detection of features at or outside the outer edges of galaxies may be used
to obtain an estimate of the rate of current and recent interactions, the merger rate, 
the frequency and importance of external gas accretion from the intergalactic reservoir,
the number of companion galaxies, statistics on the asymmetries of galactic haloes, 
and the disc structure overall. The intrinsic limitations in visual detections
include naturally the depth and spatial resolution of the data, the flat-fielding
accuracy, and the effects of interference from other perturbing astronomical or
instrumental sources, such as scattered light from nearby bright stars and image
artefacts, and the techniques used to look at the data (including the visual
acuity of the person performing the detection, of course!). Recent work on detecting 
faint features outside the main galaxy discs include those by \citet{delgado10},
\citet{tal10}, \citet{adams12}, and \citet*{atkinson13}. On the other hand, a morphological 
classification of mostly bright inner features within the discs of an initial set of 
galaxies from the Spitzer Survey of Stellar Structure in Galaxies (S$^{4}$G; \citeauthor{sheth10} 
\citeyear{sheth10}) was made by \citet{buta10}, with
classifications for the remaining galaxies in \citet{buta14}. An attempt to classify 
tidal features in S$^{4}$G galaxies, including
shells, was made by \citet{kim12}. Other major attempts to visually detect inner features in fairly 
large samples of nearby galaxies include those by \citet{fukugita07} and \citet{nair10}.

In future the number of suitably observed galaxies will increase to millions 
(e.g., with the Large Synoptic Survey Telescope, LSST), and it will not be feasible to 
perform human eye based feature detection by experts in these new samples. Therefore, visual search 
for faint features in relatively large galaxy samples, consisting of thousands of galaxies, 
such as S$^{4}$G, will also form a good training basis for automated computer algorithms 
that recognize patterns and classify them in the future. We have selected the visual
detection method in this work because we are just beginning to look for tidal and
other types of outer features. Quantitative or automatic methods are already good in quantifying
something that is known to exist in high S/N data, and will grow increasingly powerful in
detecting faint features in images in the future \citep[c.f.][]{hales12}. Our current 
effort emphasizes the detection and discovery of subtle new features, possibly related 
to tidal interaction or accretion, which are best picked up by eye, but can perhaps
be automatically detected with sophisticated codes in the future. Follow-up work may 
be able to quantify our new discoveries. In this paper we refer to already performed 
quantitative work that was based on the high S/N regions of the galaxies 
in the S$^{4}$G sample \citep{kim12,zaritsky13,holwerda14}, and therefore, inside $R_{25}$. 
The current paper thus complements the earlier work
on S$^{4}$G galaxies and extends it farther out in radius, where it presents discoveries of
faint features that should be quantified in the future when higher S/N observations are
available. Ellipse fits to the {\it Spitzer}/IRAC images and parameters derived from 
these fits are given in \citet{munoz14} and are available
in the NASA/IPAC Infrared Science Archive (IRSA) at http://irsa.ipac.caltech.edu/data/SPITZER/S4G/.
However, it should be noted that in the outermost galaxy regions that we are surveying 
in this paper, the ellipse fits are too uncertain to be trusted and some of the outer features
cannot be approximated by ellipse fits at all, leaving the visual detection as the only viable way 
to find new faint features there. 

S$^{4}$G consists of near-infrared images and thus has some unique advantages
over conventional visual band images. First, the spectral energy distribution
of late type stars, including many luminous asymptotic giant branch stars,
peaks in the near-infrared, and may thus reveal features that are not clearly
visible at shorter wavelengths. Second, in general, the dominant light in the
near-infrared is coming from older stars than the light at shorter wavelengths,
thus revealing longer lived, major dynamical features, as opposed to recent
bursts of star formation. \citet*{eskew12}, \citet{meidt12a} and \citet{meidt12b}
show that it is possible to separate the contributions from the various stellar components
and measure the mass directly with the help of S$^{4}$G near-infrared images. Therefore, 
the longer-term time evolution of galaxies can be better studied. Third, the effects 
of cold dust that can block features from view is dramatically reduced in the near-infrared. 
In our study, in which we look at features mostly outside the main galaxy bodies or 
at the edges of them, the effects of dust are generally thought to be less important 
than closer to the centre of galaxies, but some of the features that we classify,
such as shells, polar rings or even warps, may be blocked from
view at least partially at visible light wavelengths. Additional benefits of 
S$^{4}$G, as explained in Section~$\ref{data}$, are the uniformity and depth 
of the S$^{4}$G images across the sample and finally, the spatial coverage of 
the images, which around most sample galaxies extends to at least 1.5 $\times$ $R_{\rm 25}$ in
radius, making this sample amenable to morphological classification of faint 
features in the outer parts of galaxies.

\section{Sample and Data}
\label{data}

The sample we used is the full S$^{4}$G sample \citep{sheth10}, consisting of 2,352 galaxies 
(ten of the 2,331 galaxies specified in \citeauthor{sheth10} \citeyear{sheth10} were not 
observed, mostly because they were close to a very bright star, and 31 galaxies were added) with 
systemic velocity $V_{sys,radio}$ $<$ 3000~km~s$^{-1}$, corresponding to a distance $d$
$<$ 45 Mpc for a Planck mission based Hubble constant \citep{planck14} of 67~km~s$^{-1}$~Mpc$^{-1}$
and a distance $d$ $<$ 41 Mpc for a Hubble constant of 71~km~s$^{-1}$~Mpc$^{-1}$, total 
corrected blue magnitude $m_{B\rm corr}$ $<$ 15.5, blue light isophotal angular diameter 
$D_{25}$~$>$~1$\farcm 0$, and a Galactic latitude $|b|$~$>$~30$\degr$ \citep{sheth10}. 
All the galaxies in this sample were imaged with the {\it
Spitzer Space Telescope}'s Infrared Array Camera \citep[IRAC;][]{fazio04}. We used the
channel 1 (3.6 $\mu$m) mosaics made of eight 30 second frames per spatial position. The survey 
is described in detail in \citet{sheth10} which is the main reference for the S$^{4}$G sample
and data. 597 galaxies in the sample already had observations in the {\it Spitzer} Heritage
Archive, and almost all of them have a total frame time depth of at least 240 seconds (that of the
new observations). The only
exceptions are NGC 5457 (96s), NGC 0470 and NGC 0474 (150s), and NGC 5218, NGC 5216, and NGC 5576 
(192s). Several of the archival observations are from the {\it Spitzer} SINGS \citep{kennicutt03} 
and LVL \citep{dale09} Legacy Projects that had a very similar mapping strategy to the S$^{4}$G 
observations.

We started with the basic calibrated data (BCDs) that are the fundamental IRAC
pipeline-reduced images from the individual exposures. The data were subsequently
run through the S$^{4}$G Pipeline 1 that mosaics them together using the Space
Telescope Science Data Analysis System ({\sc STSDAS}) dither package (\citeauthor{sheth10}
\citeyear{sheth10}; Regan et al., in preparation). Cosmic rays are
eliminated in this process and the images are drizzled together to a mosaic that
has $0\farcs 75$ pixels (the original pixel size is about $1\farcs 2$).
\citet{sheth10} give more details on pipeline processing. We used only the
3.6 $\mu$m mosaics to search for faint outer features. The 4.5 $\mu$m images are 
usually almost identical to the 3.6 $\mu$m images, but are farther from the peak 
of the old stellar population spectral energy distribution, and may suffer from 
hot dust contribution. It should be noted that the 3.6 $\mu$m band contains the 
3.3 $\mu$m PAH emission band, while the 4.5 $\mu$m band contains a CO absorption
band.

\section{Detection and Classification Methodology}
\label{methodology}

We displayed each galaxy with the {\sc SAOImage DS9} astronomical imaging and data
visualization application \citep{joye03}, using both the histogram equalization
and log scales, and in both black-and-white and rainbow color schemes, adjusting
to the extremes of contrast and sampling carefully the contrast in between the 
extremes. We also experimented with making unsharp-masked versions of the images,
but did not use those in the final classification of the outermost features, as they did
not help in the detection of the outermost features. 

We can reliably classify features down to a per pixel surface brightness 
level of 0.03 MJy/sr (21.5 Vega mag~arcsec$^{-2}$, 24.3 AB mag~arcsec$^{-2}$, or about
2.5~$\sigma$ above the background level) 
based on the faintest detected structures (the polar ring candidates) and assess 
asymmetries in the outer isophotes at about 0.01 MJy/sr (22.7 Vega mag~arcsec$^{-2}$
or 25.5 AB mag~arcsec$^{-2}$) level at 3.6~$\mu$m. 

\begin{figure}
\centering
\includegraphics[width=8cm]{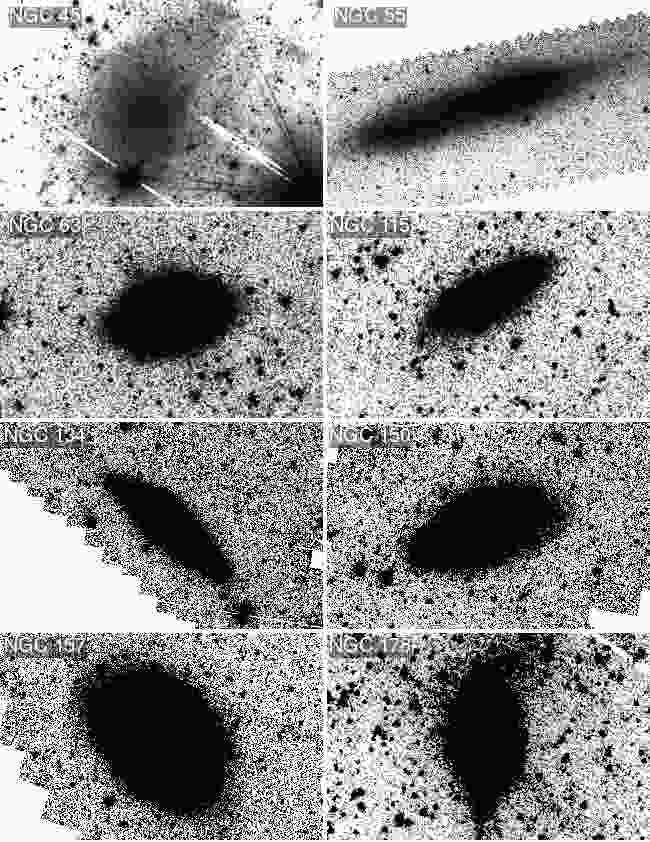}
\caption{Images of asymmetric outer discs in the S$^{4}$G sample. 
Images of all detected asymmetric outer discs are available in the online version of the Journal.\label{fig1}}
\end{figure}

\begin{figure}
\centering
\includegraphics[width=8cm]{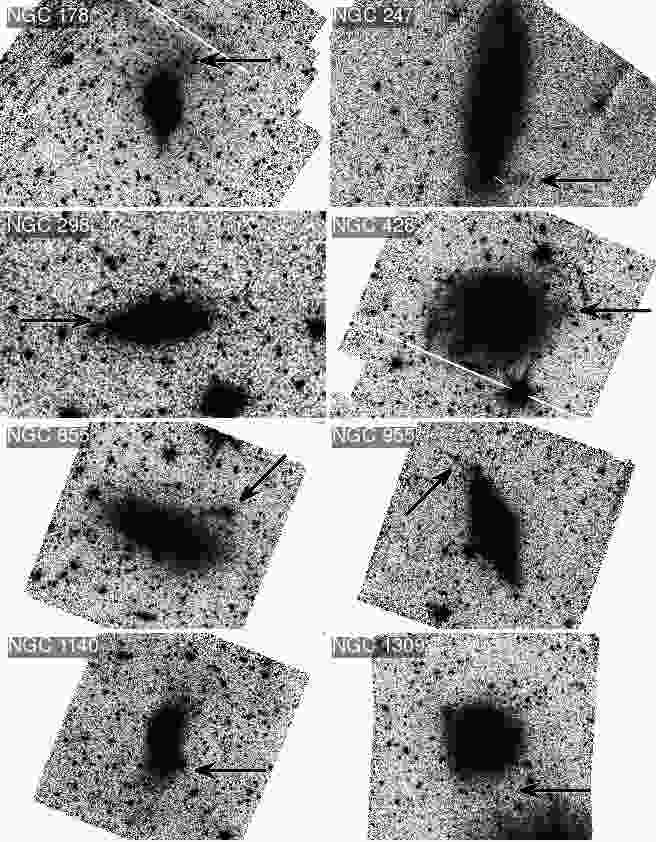}
\caption{Images of extensions in the S$^{4}$G sample. Images of all detected
extensions are available in the online version of the Journal.\label{fig2}} 
\end{figure}

\begin{figure}
\centering
\includegraphics[width=8cm]{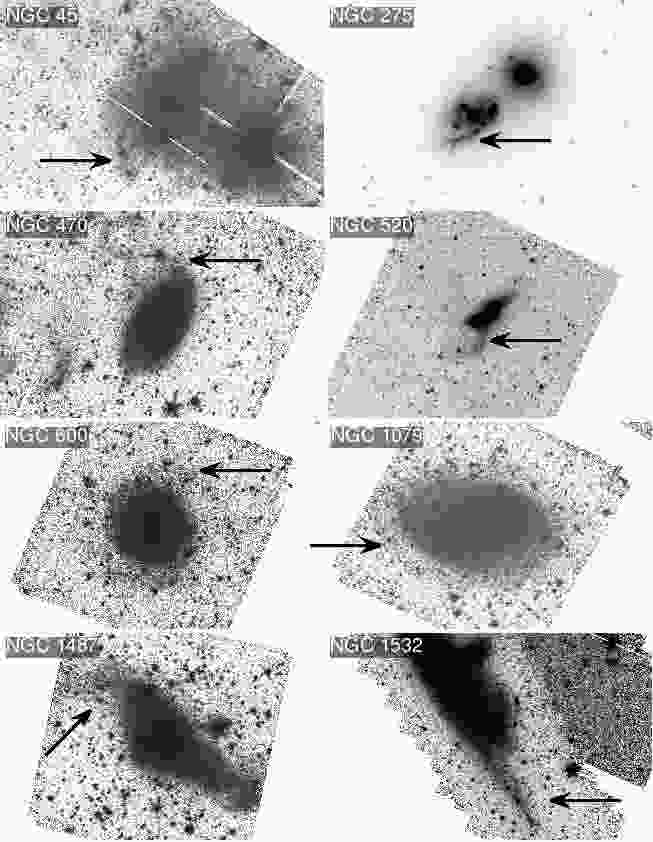}
\caption{Images of tidal tails in the S$^{4}$G sample. Images of all detected
tidal tails are available in the online version of the Journal.\label{fig3}} 
\end{figure}

\begin{figure}
\centering
\includegraphics[width=8cm]{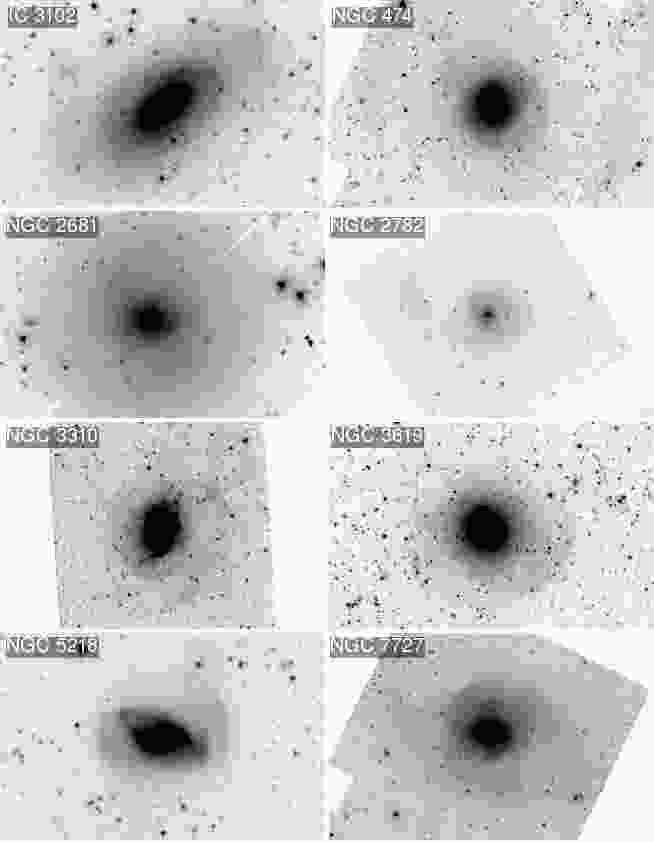}
\caption{Images of shells in the S$^{4}$G sample.\label{fig4}} 
\end{figure}

\begin{figure}
\centering
\includegraphics[width=8cm]{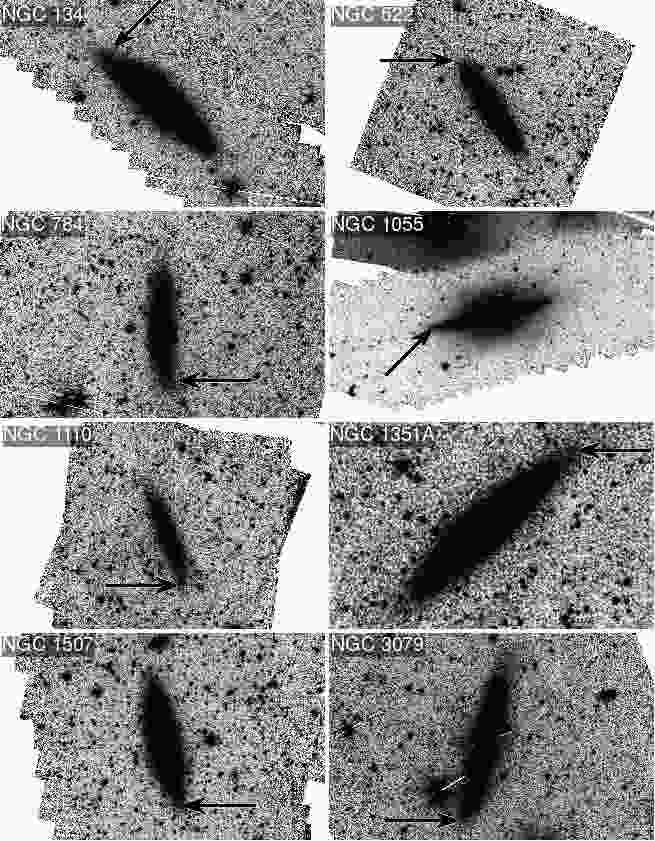}
\caption{Images of warps in the S$^{4}$G sample. Images of all detected warps are 
available in the online version of the Journal.\label{fig5}} 
\end{figure}

\begin{figure}
\centering
\includegraphics[width=8cm]{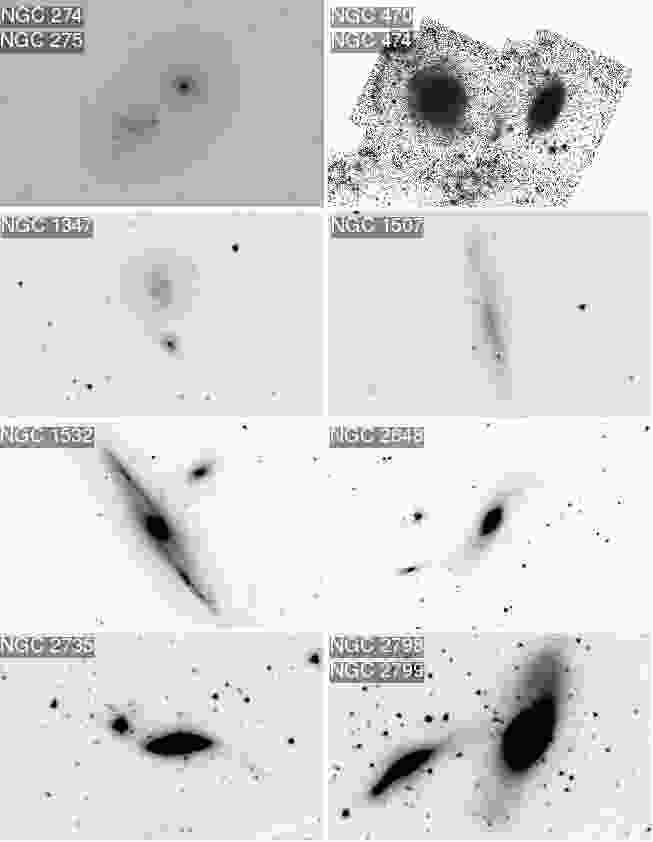}
\caption{Images of interacting galaxies in the S$^{4}$G sample. 
Images of all detected interactions are available in the online version of the 
Journal.\label{fig6}} 
\end{figure}

\begin{figure}
\centering
\includegraphics[width=8cm]{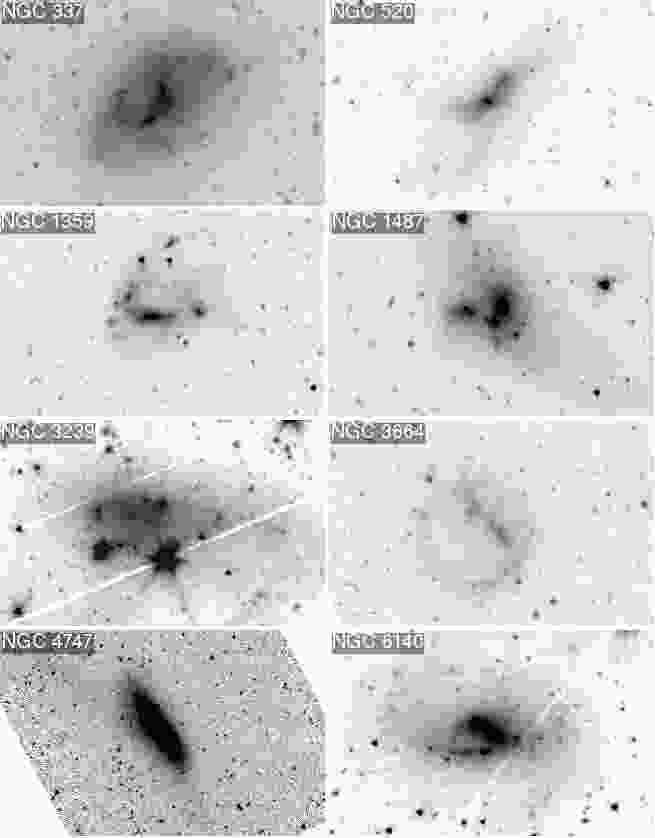}
\caption{Images of merging galaxies in the S$^{4}$G sample. 
Images of all detected mergers are available in the online version of the 
Journal.\label{fig7}} 
\end{figure}

\begin{figure}
\centering
\includegraphics[width=8cm]{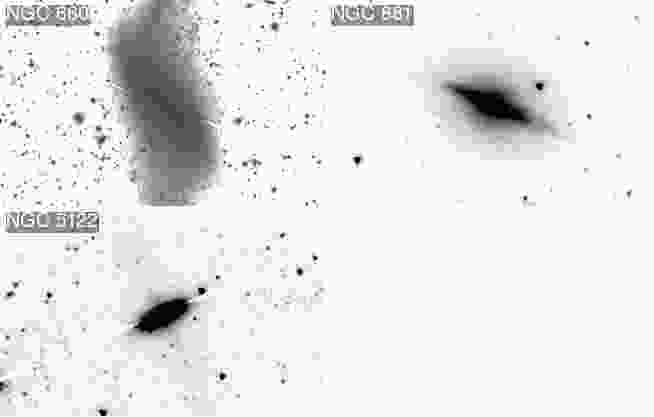}
\caption{Images of galaxies with polar rings in the S$^{4}$G sample.\label{fig8}} 
\end{figure}

The first author of this paper looked at every galaxy, and five of the coauthors 
of this paper looked at a few dozen to hundreds of separate galaxies each, 
so that each galaxy was checked by at least two persons. The detected 
features were iterated upon until all the classifiers that looked at any
given feature agreed. Immediate agreement was found for more than 2/3 of the features. 
The whole team of authors of this paper discussed the detected features, and a
consensus was formed on the discovered features reported in this paper.

We searched for eight kinds of features (in no case was the same feature 
classified as belonging to two or more different classes listed below, and
every feature was classified as belonging uniquely to one of the following
classes):

\begin{enumerate}

\item Asymmetries of the {\it outer} isophotes. If the outermost visible
isophotes were not elliptical, we called the galaxy `asymmetric.' Irregular
outer isophotes were not a sufficient reason to classify a galaxy asymmetric if the 
overall outer isophote appearance was elliptical. We did not have any
cases of symmetric boxy isophotes that we would call an asymmetry by our rule. However, if the 
nucleus was offset or the inner parts were lopsided, but the outermost visible 
isophotes were smooth and elliptical, we did not call the galaxy asymmetric. 
Lopsidedness, based on the inner high S/N parts of a small 
subsample of 167 S$^{4}$G galaxies is discussed in another paper \citep{zaritsky13}. 
All of the discovered outer disc asymmetries, as well as all the other features that we 
detected and classified, are given in Figures~\ref{fig1} -- \ref{fig8} and they are 
tabulated in Table~\ref{table1} that also shows the T-types and the 3.6 $\mu$m absolute 
AB magnitudes.

\item A clear extension on any `side' of the galaxy. An `extension' is usually
a narrow feature extending clearly far out from the edge of the galaxy. In no 
case was the same feature called both an `asymmetry' and an `extension.' 
Most extensions do not appear to be associated with spiral arms, but
in a few cases spiral arms extend well outside the visible disc or main body of
the galaxy, and were thus classified as `extensions.'

\item Warps of the disc galaxies (for more on edge-on
galaxies in the S$^{4}$G sample see \citeauthor{comeron11} \citeyear{comeron11}
and \citeauthor{comeron12} \citeyear{comeron12}).
These were only looked for in galaxies that were of very high
ellipticity as seen by eye. We looked for visually discernible
deviations from a straight line on both sides of the centre of a galaxy and
called the galaxy warped if either side (or both) showed a visually detectable
curvature. We did not pay attention to the derived inclination values of the 
galaxies while searching for warps. Therefore, some of the elongated nearly
face-on galaxies were classified as `warped.' In Section 5 we present 
statistics only for truly warped galaxies (high inclination galaxies), but keep
all the original warp classifications in Table~\ref{table1}. Warps are different 
from asymmetries in truly inclined galaxies, as a warped galaxy may still have a 
symmetric disc in its plane, but the plane itself is twisted.

\item Tidal tails. These are curved features outside the main bodies of
the galaxies, but connecting to them in most cases. These features are usually
found in galaxies that appear to be interacting or merging, but sometimes
tidal-like features are seen in apparently non-interacting galaxies, and could
result from past mergers. These features extend and curve around galaxies for
much longer distances than extensions (usually comparable in length to half the
catalogued galaxy diameter). In some cases a curvy shorter feature outside the
main galaxy disc was also called a tidal tail. It is possible that some of the
features classified as tidal tails are associated with outer ring features such
as in NGC~1079 \citep{buta95}.

\item Shell features. Definitions of shell features are given in \citet{lia85}.
These features exist usually around elliptical and lenticular galaxies 
\citep[e.g.,][]{malin80,schweizer88}. They usually have sharp, curved edges in their
light distribution at their outer edge, with the concave part always pointing
towards the galaxy centre. There is no reason why they should exist only around
elliptical galaxies, and therefore, we searched for them in disc
galaxies as well. However, there is some expectation that shells should be less
likely in disc galaxies, because they arise from deeply plunging orbits that would
perturb the disc. In Section~\ref{statistics} we discuss the differences
of our detected shell features from those of \citet{kim12} in the S$^{4}$G sample.

\item Interacting and merging galaxies. The {\it revealing} and defining sign of an ongoing
interaction between disc-like galaxies is a bridge or some connecting material 
between two galaxies. However, early-type galaxies may not have such obvious
signs of interaction and may be missed in a visual search. We did not use any
velocity information in our interaction/merger classifications. Thus, two or more galaxies
close to each other in systemic velocity but with no bridge feature between them
would be classified as `companions' (see below). A merger leaves
behind a very disturbed morphology and tidal features, but no signs of two
(or more) separate galaxy bodies are left over. Kinematic observations of
galaxies in our merger class should be able to reveal whether they are truly
mergers or just irregular galaxies.

\item Polar rings around main galaxy bodies. For definitions, see again
\citet{lia85}. Polar rings are features that are usually perpendicular to the
position angle of the major axis of the galaxy, but they can exist at other
apparent angles as well. They are often needle-like features extending outside
the main bodies at a sharp angle to the major axis.

\item Companion galaxies. We looked for nearby companion galaxies in the imaged
area around the sample galaxies and if found, checked their systemic velocity. If
the systemic velocity, checked using the NASA/IPAC Extragalactic Database (NED), was 
within $\pm$600~km~s$^{-1}$ of the
target galaxy systemic velocity, we classified it as a companion galaxy. Previous
studies using isolation criteria or searches for companion galaxies often use systemic
velocity ranges between 500~km~s$^{-1}$ and 1000~km~s$^{-1}$ \citep[e.g.,][]{zaritsky93,sales05}.
Our value of 600~km~s$^{-1}$ is a compromise between these two, and represents the 
velocity dispersion of a modest size cluster of galaxies (less than Virgo), and more 
than that of a galaxy group, so we would find galaxy group members. There can be 
interacting galaxies beyond this cut in some rare cases, such as possibly
NGC~4435 and NGC~4438. Note that the physical size of the imaged area varied a lot, 
as the sample galaxies are at distances from 1 to about 60 Mpc (almost all of them within about
40 Mpc), and vary in physical size as well. If a velocity measurement was not available 
for a nearby galaxy, we did not include it as a physical companion galaxy.

\end{enumerate} 

\begin{figure*}
\centering
\includegraphics[width=12cm]{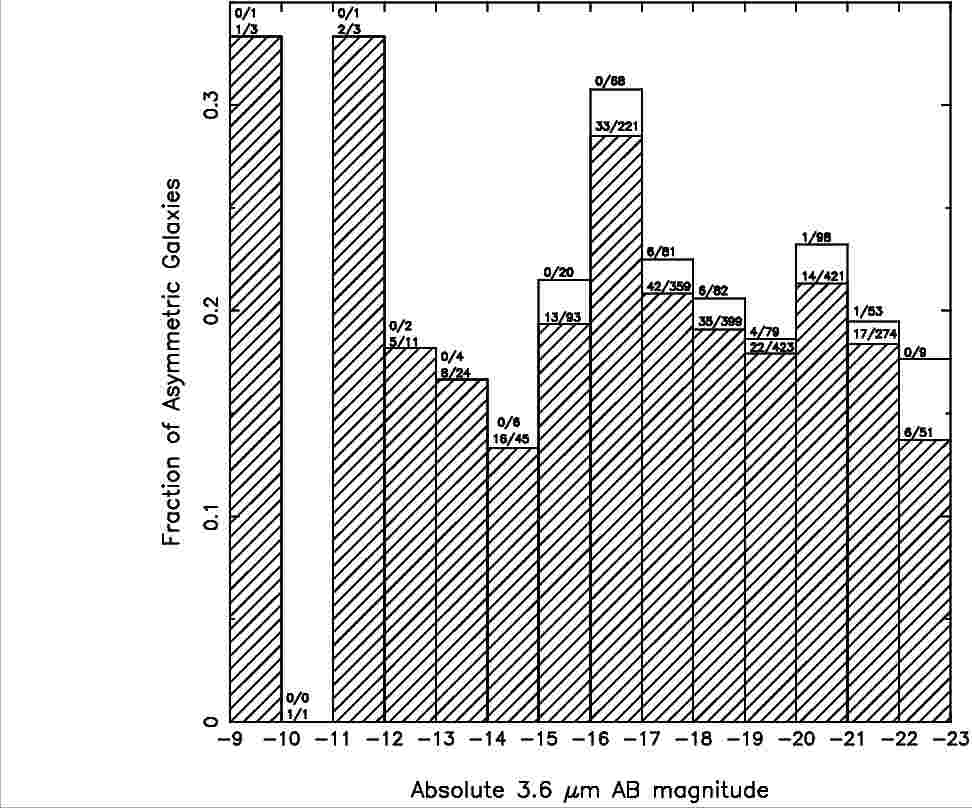}
\caption{Fraction of asymmetric galaxies as a function of 3.6 $\mu$m absolute AB magnitude. 
The fraction of uncertain (marked by the symbol `?' in Table~\ref{table1}) asymmetries is plotted without hatching.
The fractions of 1) asymmetric galaxies in a given magnitude bin with the overall uncertainty flag 'U' in
Table~\ref{table1} over all asymmetric galaxies in the given magnitude bin
(see Table~\ref{table1}) and 2) all galaxies in a given magnitude bin with the uncertainty flag 
`U' over all galaxies in that magnitude bin are given above the corresponding magnitude column. 
The luminosity could not be determined for four asymmetric galaxies and 21 galaxies in the whole sample.\label{fig9}} 
\end{figure*}

\begin{figure*}
\centering
\includegraphics[width=12cm]{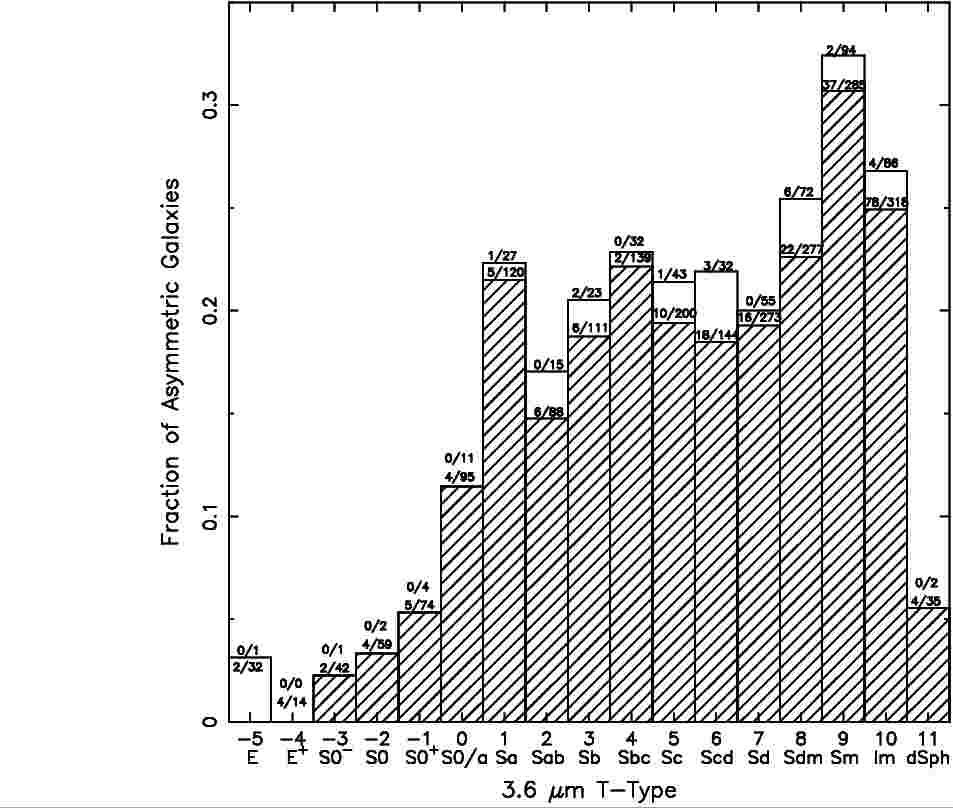}
\caption{Fraction of asymmetric galaxies as a function of 3.6 $\mu$m T-type. The fraction 
of uncertain (marked by the symbol `?' in Table~\ref{table1}) asymmetry detections is plotted without hatching. 
The fraction of asymmetric galaxies in a given T-type bin with the overall uncertainty flag 'U' in
Table~\ref{table1} over all asymmetric galaxies in that T-type bin (see Table~\ref{table1}) 
and the fraction of all galaxies in a given T-type bin with the uncertainty flag `U' 
are given above the corresponding T-type column. The T-type could not be determined 
for six asymmetric galaxies and 18 galaxies in the whole sample. The ``dSph'' type
includes dE, dS and Sph types.\label{fig10}} 
\end{figure*}

\begin{figure*}
\centering
\includegraphics[width=12cm]{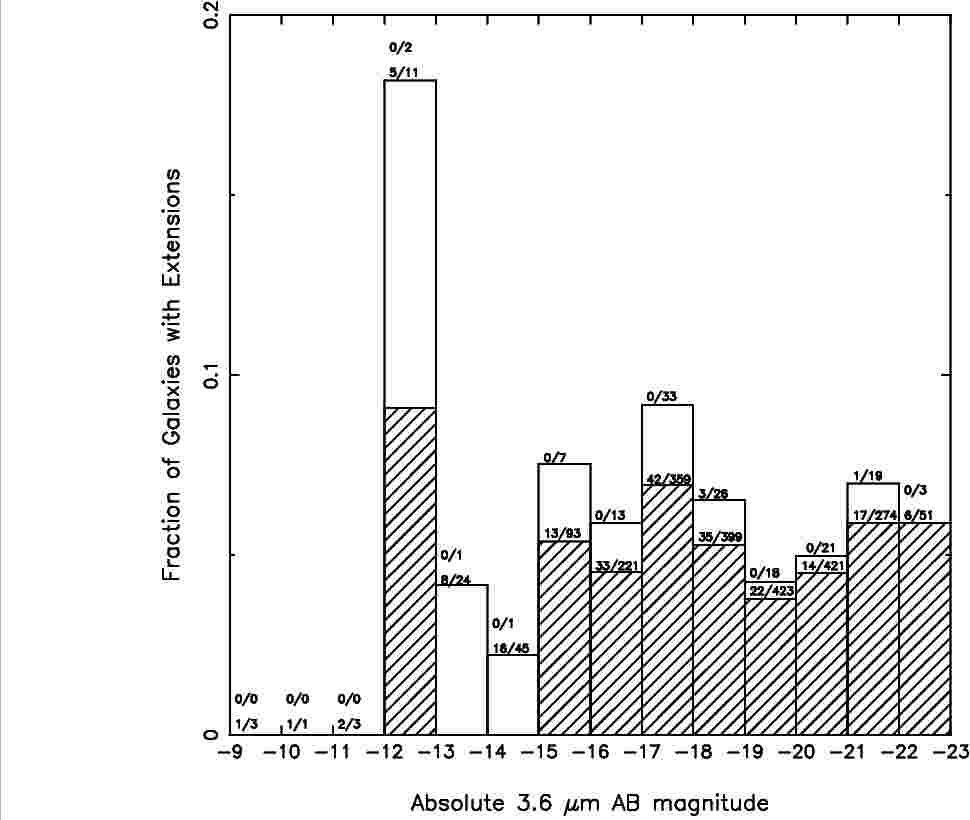}
\caption{Fraction of galaxies with extensions as a function of 3.6 $\mu$m absolute AB magnitude. 
The fraction of uncertain (marked by the symbol `?' in Table~\ref{table1}) extension detections is plotted without hatching.
The fractions of galaxies with extensions in a given magnitude bin with the overall uncertainty flag 'U' in
Table~\ref{table1} over all the galaxies with extensions in the given magnitude bin 
(see Table~\ref{table1}) and the fraction of all galaxies in a given magnitude bin with the 
uncertainty flag `U' are given above the corresponding magnitude column. The luminosity 
could not be determined for one galaxy with an extension and 21 galaxies in the whole 
sample.\label{fig11}} 
\end{figure*}

\begin{figure*}
\centering
\includegraphics[width=10cm]{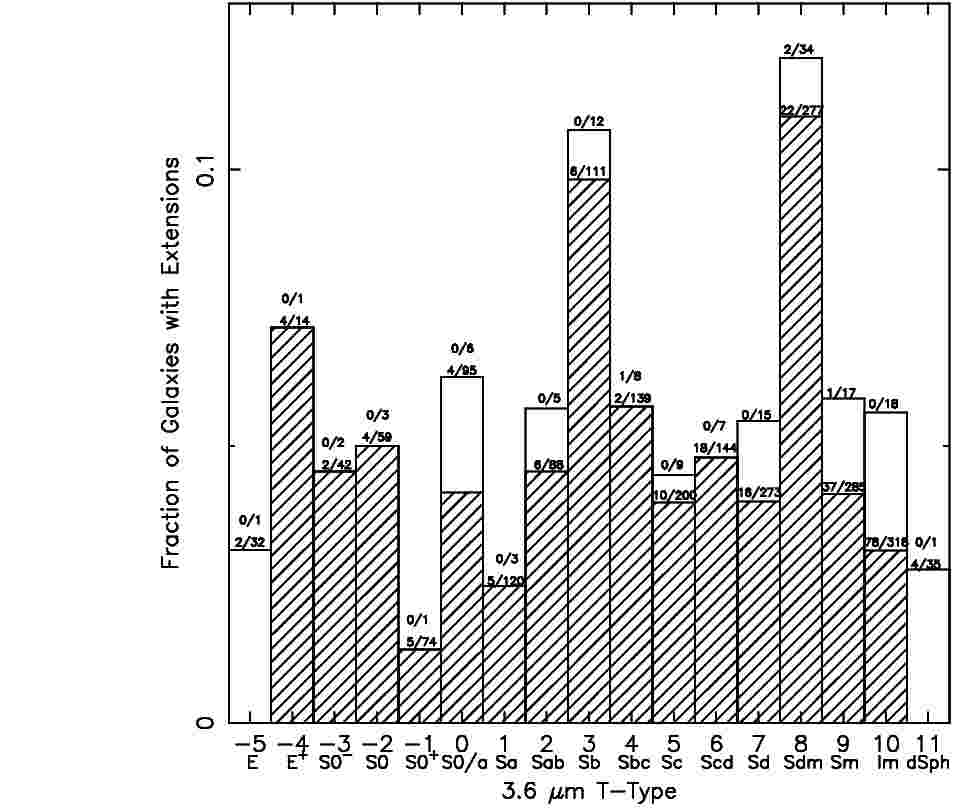}
\caption{Fraction of galaxies with extensions as a function of 3.6 $\mu$m T-type. The fraction 
of uncertain (marked by the symbol `?' in Table~\ref{table1}) extension detections is plotted without hatching. The fraction of 
galaxies with extensions in a given T-type bin with the overall uncertainty flag 'U' in
Table~\ref{table1} over all the galaxies with extensions in the given T-type bin (see Table~\ref{table1}) 
and the fraction of all galaxies in a given T-type bin with the uncertainty flag `U' are 
given above the corresponding T-type column. The T-type could not be determined for 
two galaxies with extensions and 18 galaxies in the whole sample. The ``dSph'' type
includes dE, dS and Sph types.\label{fig12}} 
\end{figure*}

In addition to these detected features, we kept track of image features that
made the search for these faint outer features associated with the galaxy 
very uncertain or impossible. Such interfering features include 
most often bright stars that are located on top of, or near the edges of the
main bodies of the galaxies, remaining image artefacts, pointings where the galaxy
is near the edge of the field of view, and galaxies with very patchy and faint 
morphology, showing no continuous main bodies. All these galaxies are marked with a 
`U' in Table~\ref{table1}. Note that `?' in the Classification column in 
Table~\ref{table1} after a feature symbol means that the assignment of a feature
into one of the above-mentioned classes was uncertain, but not due to the  
interfering feature uncertainty marked by the letter `U.'

The displayed asymmetries most often have a clear departure from a pure elliptical
outer edge. This can in some cases be due to the presence of strong spiral
arms that continue outside the main body of the galaxy (e.g., NGC 2750 in
Figure~\ref{fig1}). Some galaxies are not elliptical at all in their outer
regions, such as NGC~3628 in Fig.~\ref{fig1}. More extreme departures from
elliptical outer isophotes are also seen.

There may be some overlap between the categories of `asymmetric' and
`extensions.' However, we considered an extension to be a feature clearly
protruding out of the galaxy disc, instead of, for example, a slight extension
of one `side' of the galaxy compared to others. Figure~\ref{fig2} illustrates
what we consider to be extensions. Often the extensions are small-scale
features protruding out from the galaxy.

There is possibly some overlap between the `extension' and `tidal tail' 
categories as well. We considered extensions usually to be linear features 
protruding out of the galaxy discs at close to right angles to the major axis, 
whereas curved features often seen starting close to the end of the major axis 
are considered to be tidal tails. The presence of an `interacting' or `merger' 
morphology is a reason to consider an extended feature a tidal tail instead 
of an extension.

Real warps of the disc are naturally seen only in fairly high inclination 
galaxies ($i$ $>$  $\sim$ 65$\degr$; Fig.~\ref{fig5}). But as stated earlier,
we allowed elongated and twisted irregular galaxies to have
`warps' as well. However, when calculating the statistical numbers of
warps, we only counted warps in highly inclined galaxies as explained in
detail below.

Polar rings were the hardest features to discern in the images, and they are
rare. Image artefacts, such as, for example, the presence of column pulldown 
(examples are seen in the IRAC Instrument Handbook\footnote{http://irsa.ipac.caltech.edu/data/SPITZER/docs/irac/ 
\\iracinstrumenthandbook/}),
may conspire to create a polar-ring -like impression. Partly for this reason, our
polar ring assignments are all uncertain.

\section{The Catalogue, Statistics, And Correlations of Features}
\label{statistics}

The main catalogue is presented in Table~\ref{table1}. We list the galaxy name,
the 3.6 $\mu$m T-type (\citeauthor{buta10} \citeyear{buta10}; \citeauthor{buta14}
\citeyear{buta14}), the galaxy absolute 3.6 $\mu$m
AB magnitude \citep{munoz14}, and the presence of any 
detected features. The 17 Hubble types from E to dE/dS/Sph are assigned numerical T
values as follows: $-5$ (E), $-4$ (E$^{+}$), $-3$ (S0$^{-}$), $-2$ (S0$^{0}$), $-1$
(S0$^{+}$), 0 (S0/a), 1 (Sa), 2 (Sab), 3 (Sb), 4 (Sbc), 5 (Sc), 6
(Scd), 7 (Sd), 8 (Sdm), 9 (Sm), 10 (Im), and 11 (dE, dS or Sph) \citep*{vaucouleurs64,binggeli85,kormendy12a,kormendy12b}. 
The AB magnitudes were calculated using the mean redshift-independent 
distance from NED whenever available, and a Hubble constant of 71~km~s$^{-1}$~Mpc$^{-1}$ 
otherwise.

\subsection{Asymmetric Galaxies and Extensions}
\label{asymstats}

Asymmetries are by far the most common feature we found. We found 506 asymmetric
galaxies in the sample, or 22$\pm$1 per cent (uncertainty is purely statistical, calculated
as the standard deviation of a binomial distribution). There is of course a large range in the
magnitude of asymmetry. Even if we remove the questionable cases, we are left
with 469 or 20$\pm$1 per cent asymmetric galaxies in a sample of 2,352 galaxies.
Earlier studies, such as \citet{rix95} who examined a sample of 18 face-on spiral
galaxies in the $K'$-band and found that a third of them were lopsided, and 
\citet[][and see references therein]{reichard08}, who inspected over 25,000 
galaxies in the Sloan Digital Sky Survey, measured lopsidedness in the 
whole galaxy disc, although the latter study found that lopsidedness (that causes 
asymmetry) increases with radius \citep[see also][]{zaritsky13}. Our result for the 
asymmetry fraction can also be compared to the quantitative morphological classification of
S$^{4}$G galaxies inside their luminous bodies by
\citet{holwerda14}. They find that roughly one quarter of the S$^{4}$G sample galaxies are
`disturbed,' meaning that according to CAS criteria \citep{conselice03} these
galaxies have an asymmetry value larger than their smoothness value, and the 
absolute value of the asymmetry is greater than 0.38. Note that the
quantitative criterion uses all the pixels inside the $\sim$~$R_{25}$ radius but not outside
of it, in the regions that we are concerned with in this paper and thus there may
be a detection of asymmetry in the quantitative classification scheme but
not in our visual examination which only considered the radii at $R_{25}$ and
outside of it. Therefore, the asymmetry fractions in the current paper and in \citet{holwerda14} 
are not directly comparable.

We plot the distributions of absolute 3.6 $\mu$m AB magnitudes and T-types for
asymmetric galaxies in Figures~\ref{fig9} and \ref{fig10}, respectively. As less 
massive late-type galaxies are structurally more prone to outer disc
disturbances than early type massive elliptical and lenticular galaxies due to 
the often younger and kinematically more uniform stellar populations of the less massive,
late-type galaxies, we
expect, and observe, that the fraction of asymmetries goes up towards the later
types. Similarly, using quantitative measurements in a small S$^{4}$G subsample,
\citet{zaritsky13} found that there is greater lopsidedness for galaxies of later 
type and lower surface brightness. A similar increase in asymmetry towards later 
types was seen in the quantitative S$^{4}$G morphology paper \citep{holwerda14}. 
Also \citet{bridge10} found that the least massive galaxies have a higher merger
rate and therefore presumably look more asymmetric than the massive galaxies at 
low redshifts (their sample went down to $z = 0.2$), consistent with our results at 
$z$ close to zero. The S$^{4}$G sample is dominated by `extreme late-type galaxies' (Sd, Sdm, Sm,
and Im), and these types are characteristically asymmetric, especially Sdm and Sm. 
The use of a radio radial velocity in the sample definition weighted the sample towards 
these types. The luminosity distribution shows that the S$^{4}$G sample is 
magnitude-limited, and therefore distant, intrinsically faint galaxies are not 
included. 

It is notable that only 14/48 or 29$\pm$7 per cent of the galaxies classified as
undergoing an interaction in our study have asymmetric outer discs (Table~\ref{table2}). 
On the other hand, out of the 506 galaxies with asymmetric outer disc classifications in our
study, only 14 are interacting galaxies, based on visual bridges
between galaxies. This result, combined with the fact that only 64/212 or
30$\pm$3 per cent of galaxies that have companions within the mapped area have
asymmetric outer discs (Table~\ref{table2}), has been used as evidence by \citet{zaritsky13} to argue
that small lopsidedness in the S$^{4}$G sample of galaxies is mainly caused by
internal factors, such as dark halo asymmetries. Small asymmetries in the dark halo 
can give rise to larger, observable stellar asymmetries \citep{jog09}. However, one should keep in
mind that our low correlation of asymmetries with interactions and 
companions may be caused by 1) the outer isophotes remaining relatively 
elliptical even if there is a minor perturbation (satellite interaction
or minor merger); 2) not being able to count all the companions because of
limitations in the field of view and depth of the survey; and 3) as compared
to such samples as the Galaxy Zoo \citep{casteels13}, our sample size being
still relatively small.

There may be several different origins for the outer disc asymmetries, including
internal effects, such as lopsidedness \citep{zaritsky13}. If we assume that 
they are overwhelmingly due to interactions, it is possible to make an argument
for the duration of asymmetries in outer discs due to interactions as follows.
The magnitude of the brightness asymmetries in the outer halves of the discs of 
several galaxies studied here is around 50 per cent. We estimate whether this is high 
enough to produce significant torques and radial flows if the disc mass follows the
disc light. We consider for simplicity a $180^\circ$ asymmetry of magnitude
$A\sim50$ per cent. Then the outer disc mass in four quadrants varies in azimuth as
$M_4(1-A/2)$, $M_4$, $M_4(1+A/2)$, and $M_4$, where $M_4$ is one-quarter of
the disc mass in the radial interval of the asymmetry, say, the outer half of
the disc.  In this case, the forward and backward torques on the minor axes
of the asymmetry are

\begin{equation}
{\rm Torque} = G M_4 M_4 [ (1+A/2) - (1-A/2) ] R / (1.4 R)^2
\end{equation}
where $R$ is the average radius of the outer disc and $2^{0.5}R\simeq1.4R$ is the
distance between quadrants in the outer disc where the masses are $M_4$ and
$M_4(1\pm A/2)$.

Setting the torque on a quadrant equal to the time derivative of $M_{4}RV$ 
(where $V$ is the rotation speed and $R$ is the radius), and considering that the mass 
and rotation speed do not change much during the subsequent adjustment, we obtain
the radial outflow or inflow speed $V_R$ that is driven by this asymmetry:

\begin{equation}
V_R/V = {{ A M_{\rm outer\;disc} } \over { 8 M_{\rm gal}}},
\end{equation}
where $M_{\rm outer\;disc}=4M_4$ is the total disc mass in the radial range
of the asymmetry and $M_{\rm gal}$ is the total galaxy mass inside the radius
$R$ that gives the rotation speed, using the equation $V^2=GM_{\rm gal}/R$.

If we consider that the outer disc mass is $\sim10$ per cent of the total mass inside
the radius $R$, then $V_R/V\sim0.012A$, which is a fairly small effect at any
one time, giving, e.g., $V_R\sim 1$ km s$^{-1}$.   This is barely enough to
relax and mix an outer asymmetry spanning a radial range of $\sim10$ kpc by
disc torques in a Hubble time.

More important would be mixing and smearing of the asymmetry by shear given
the rotation curve from dark matter. In the outer disc, the rotation time is
$2\pi R/V\sim0.5$ Gyr, so this would be the approximate lifetime of an
initially $180^\circ$ asymmetry at $R\sim20$ kpc before shear turns it into a
spiral or tidal arm. Galaxy interactions that perturb discs down to $R\sim5$
kpc would produce tidal arms four times faster, in $\sim0.1$ Gyr.

However, asymmetries in the far outer disc might still be visible after several
rotations, or some $\sim4$ Gyrs, as suggested by the cosmological zoom re-simulations in
Section~\ref{simulations}. If only half the galaxies have a strong interaction
which leaves signs 
that last for 4~Gyrs, then in 10 Gyrs (a typical galaxy age) we would see 40 per cent
of that half with an outer disc asymmetry, or a total of 20 per cent of all galaxies
that we see at any given time would have an asymmetry, assuming that interactions
take place at random times during the 10 Gyr galaxy age. This
20 per cent is comparable to the fraction of discs in our survey that have
perceptible outer disc asymmetries, suggesting that many of these structures could
be remnants of interactions less than about $\sim4$ Gyrs ago, with the older 
structures now mixed and invisible. It should be noted that the lifetime estimates 
we get this way depend on the assumption that the asymmetries are exclusively due
to interactions. If other factors are in play, the lifetime of the features 
could be shorter or there may have been fewer interactions. 

It is also interesting to compare our estimate of about 20 per cent asymmetries to
the fractional estimate of 15 per cent of asymmetric, lopsided, warped or distorted with an 
integral-sign like appearance or tidal feature (tail, bridge or shells) in the sample 
of \citet{fernandez12} of visible light images of 466 isolated nearby galaxies with 
systemic velocities between 1,500 and 24,000 km~s$^{-1}$. This difference may be due 
to different depths of the surveys, different distance limits on the samples and a
different wavelength region surveyed.

Extensions were found in 6$\pm$1 per cent of the whole S$^{4}$G sample. Figures
\ref{fig11} and \ref{fig12} display histograms versus 3.6 $\mu$m absolute AB 
magnitudes and T-types for galaxies with extensions. We did not find any clear 
absolute magnitude  (more extensions might be expected to be seen among the
low mass, often irregular low luminosity systems) or T-type dependence for extensions.

\subsection{Warped Galaxies}
\label{warpstats}

We have found warps or possible warps in 32 edge-on galaxies among the 489
highly inclined ($i$~$>$~$65\degr$) galaxies (7$\pm$1 per cent) \citep[see also][]{comeron11}. It
is likely that the inclination has to be higher than $\simeq80{\degr}$
\citep[e.g.,][]{res98} for a warp to become visible in visual or near-IR 
observations, but we allow here for `warps' in less inclined galaxies. 
Using $80{\degr}$ as the minimum inclination, we find warps in 9 out
of 75 galaxies (12$\pm$4 per cent). Warps have been
conventionally found in H\,{\sc{i}} observations, e.g., \citet{bosma91}, where
they are often more obvious and produce higher warped fractions in the
smaller samples that were observed than here. Warp studies using visible light images 
detected a higher fraction of warps in the samples that were biased towards more
inclined, bright, late-type galaxies, e.g., \citet[][53 per cent]{sanchez03} and 
\citet[][40 per cent]{res98,res99}. The warped galaxies in our study are predominantly
less bright than $L^{*}$, and have mostly Hubble T-types from 3 
to 10 (or Sb to Im). We thus confirm the \citet{res98,res99} result concerning galaxy 
type, but we disagree with the effect of luminosities and so, presumably mass. 
We have classified ten other galaxies as warped because our classification
took place without a priori knowledge of the inclination of the galaxies. These
are mostly galaxies with low inclinations and elongated morphologies and therefore
are not `warped' in the sense of edge-on discs. For four of them, the warp detection 
is questionable.

\subsection{Tidal Tails and Interactions}
\label{tidalstats}

Tidal tails are found in 71/2,352 galaxies or in 3$\pm$1 per cent of the sample galaxies.
24 of these detections are uncertain. Because of the limited map size,
it is not clear in many cases which galaxy is causing these tidal features.
Several examples of the `diffuse' tidal tail morphology are seen (in the similar
visual classification scheme of \citeauthor{elmegreen07} \citeyear{elmegreen07} who
used it for intermediate redshift galaxies out to $z$=1.4 in the GEMS and GOODS fields).
Also some examples of the `antenna' morphology are seen. Our sample can
be used as a nearby comparison sample for future comparisons of detections of 
tidal features in higher-z galaxies.

We have recognized 31 interacting systems in the full S$^{4}$G sample used here.
These contain examples of the `M51-type' and `shrimp' galaxies as
classified by \citet{elmegreen07}. Equal mass interactions are also
represented. However, probably none of our interactions would be
classified as `assembling' in the classification scheme of \citet{elmegreen07}.
This is consistent with the common picture of galaxy evolution where
the galaxy assembly took place at high redshifts. However, among the
ten merger systems in the S$^{4}$G sample there are a few which could
be classified as `assembling' (Figure ~\ref{fig7}; NGC 337, NGC 1487). Our 
complete sample of nearby galaxies 
should again be useful in future studies that want to compare the frequency 
of galaxy assembly at higher redshifts to the current epoch (assuming the
same rest wavelength, {\it JWST} MIRI sensitivity should be sufficient to
detect similarly bright features out to a $z$ of about 0.3--0.4 and 
future detectors may be able to push this limit into even higher
redshifts where most of the galaxy assembly took place).

Interactions and mergers were also searched for in visible light images in
a parallel project \citep{knapen14}. Only 69 per cent of the
interactions/mergers in our sample are classified as such in the visible
light images. The difference can be explained by different classification
criteria used, and the larger imaged areas in the visible light images.

\begin{table*}
\caption{MORPHOLOGICAL OUTER FEATURES OF S$^{4}$G GALAXIES.\label{table1}}
\begin{tabular}{lccrlccr}
\hline
Name & T-Type & Abs.3.6 $\mu$m & Classification & Name & T-Type & Abs. 3.6 $\mu$m & Classification \\
 &  & AB Mag. &  &  &  & AB Mag. &  \\
 \hline
NGC 45    &   8  &  $-18.90$  & A,TT &      NGC 1325  &   3  &  $-20.34$  & A       \\         
NGC 55    &   9  &  $-999  $  & A &	    NGC 1326B &  10  &  $-17.73$  & A       \\	       
NGC 63    &   0  &  $-19.53$  & A &	    NGC 1332  & $-3$ &  $-21.76$  & C       \\    
NGC 115   &   7  &  $-18.68$  & A &	    NGC 1337  &   6  &  $-19.25$  & A       \\    
NGC 134   &   4  &  $-21.99$  & A?,W?,C &   NGC 1338  &   6  &  $-20.02$  & U       \\    
NGC 150   &   2  &  $-20.70$  & A? &	    NGC 1347  &   8  &  $-18.20$  & I       \\    
NGC 157   &   5  &  $-21.38$  & A &	    NGC 1351A &   5  &  $-18.93$  & W       \\    
NGC 178   &  99  &  $-18.13$  & A,E?   &    NGC 1357  &   0  &  $-21.12$  & C? 	    \\
NGC 216   &   8  &  $-18.35$  & A &	    NGC 1359  &   8  &  $-19.65$  & A,M?    \\
NGC 247   &   8  &  $-18.64$  & E,U &	    NGC 1367  &   0  &  $-21.48$  & E       \\   
NGC 254   &$-1$  &  $-19.78$  & C  &	    NGC 1385  &   8  &  $-20.12$  & E       \\   
NGC 274   &$-3$  &  $-19.45$  & I  &	    NGC 1406  &   5  &  $-20.35$  & E       \\   
NGC 275   &   8  &  $-19.42$  & TT,I &      NGC 1411  &$-2$  &  $-20.07$  & C       \\   
NGC 289   &   2  &  $-21.26$  & A,C  &      NGC 1414  &   2  &  $-17.12$  & E       \\   
NGC 298   &   8  &  $-18.03$  & E &	    NGC 1422  &   7  &  $-18.48$  & E       \\   
NGC 337   &   6  &  $-20.26$  & A,M? &      NGC 1427A &  10  &  $-17.57$  & A       \\   
NGC 337A  &   9  &  $-17.80$  & A &	    NGC 1437B &  10  &  $-17.00$  & C?      \\    
NGC 428   &   8  &  $-19.02$  & E &	    NGC 1482  &   2  &  $-20.83$  & E,C     \\
NGC 470   &   2  &  $-21.63$  & TT,I &      NGC 1484  &   8  &  $-17.98$  & E  	    \\
NGC 474   &   0  &  $-21.17$  & S,I &	    NGC 1487  &  99  &  $-17.78$  & A,TT,M? \\     
NGC 518   &$-1$  &  $-20.51$  & A &	    NGC 1495  &   8  &  $-18.86$  & A  	    \\ 
NGC 520   &  99  &  $-21.45$  & TT,M &      NGC 1507  &   9  &  $-17.99$  & W,I,C   \\
NGC 522   &   0  &  $-21.35$  & W &	    NGC 1510  &   0  &  $-16.68$  & C  	    \\
NGC 578   &   6  &  $-20.52$  & A &	    NGC 1511  &   5  &  $-20.44$  & C  	    \\
NGC 600   &   7  &  $-19.19$  & TT &	    NGC 1512  &   1  &  $-20.41$  & A,C     \\	
NGC 658   &   4  &  $-20.73$  & A &	    NGC 1518  &   9  &  $-18.01$  & A 	    \\
NGC 660   &   2  &  $-21.07$  & PR? &	    NGC 1532  &   2  &  $-22.08$  & A,TT,I  \\	
NGC 672   &   7  &  $-18.48$  & C &	    NGC 1546  &   1  &  $-20.15$  & U  	 \\
NGC 681   &   4  &  $-21.51$  & PR? &	    NGC 1553  &$-1$ & $-22.24$ & U	 \\	  
NGC 691   &   2  &  $-21.44$  & A   &	    NGC 1556  &   9 & $-18.03$ & E	 \\	  
NGC 772   &   3 & $-22.58$ & C &  	    NGC 1559  &   6 & $-20.76$ & A	 \\
NGC 784   &  10 & $-16.00$ & A,W &	    NGC 1566  &   3 & $-21.12$ & A	 \\
NGC 855   &$-3$ & $-17.52$ & E   &          NGC 1592  &  10 & $-15.77$ & A	 \\
NGC 864   &   4 & $-20.21$ & A   &          NGC 1596  &$-3$ & $-20.22$ & C	 \\
NGC 865   &   4 & $-20.14$ & A &            NGC 1602  &  10 & $-17.78$ & A,C	 \\
NGC 895   &   5 & $-20.67$ & A  &           NGC 1637  &   3 & $-19.59$ & A	 \\
NGC 908   &   3 & $-21.44$ & U  &           NGC 1679  &  10 & $-18.57$ & E	 \\
NGC 955   &$-1$ & $-20.40$ & E  &           NGC 1688  &   8 & $-18.97$ & A	 \\
NGC 986   &   2 & $-20.87$ & A  &           NGC 1808  &   1 & $-21.34$ & A	 \\
NGC 986A  &  10 & $-16.37$ & A  &           NGC 1809  &   8 & $-18.49$ & A?,E,U  \\
NGC 988   &   7 & $-19.71$ & U  &           NGC 1879  &   9 & $-18.43$ & A	 \\
NGC 1032  &$-3$ & $-21.87$ & U  &           NGC 1892  &   5 & $-18.93$ & A	 \\
NGC 1047  &  11 & $-17.76$ & A  &           NGC 2101  &  10 & $-16.85$ & A	 \\
NGC 1055  &   4 & $-21.62$ & W  &           NGC 2460  &   1 & $-21.41$ & TT	 \\
NGC 1068  &   1 & $-22.70$ & U  &           NGC 2541  &   8 & $-18.26$ & A	 \\
NGC 1079  &$-1$ & $-21.07$ & TT? &          NGC 2543  &   3 & $-20.73$ & E	 \\
NGC 1084  &   5 & $-21.41$ & A  &           NGC 2552  &   9 & $-17.54$ & A	 \\
NGC 1087  &   7 & $-20.37$ & A  &           NGC 2608  &   3 & $-20.24$ & A	 \\
NGC 1090  &   5 & $-20.84$ & A  &           NGC 2633  &   3 & $-21.03$ & A	 \\ 
NGC 1097  &   2 & $-22.81$ & A,C  &         NGC 2634A &   9 & $-18.47$ & C	 \\
NGC 1110  &   9 & $-17.33$ & W   &          NGC 2648  &   1 & $-21.41$ & A,I,C    \\
NGC 1140  &$-2$ & $-18.68$ & A,E  &         NGC 2655  &   0 & $-22.33$ & E	 \\
NGC 1179  &   6 & $-19.25$ & A   &          NGC 2681  &   0 & $-20.75$ & S?	 \\ 
NGC 1187  &   4 & $-20.84$ & C   &          NGC 2685  &$-2$ & $-19.69$ & E	 \\
NGC 1222  &$-3$ & $-20.35$ & A   &          NGC 2715  &   5 & $-20.52$ & E	\\
NGC 1249  &   9 & $-19.01$ & A   &          NGC 2731  &  99 & $-20.25$ & A	\\
NGC 1253  &   7 & $-20.17$ & C   &          NGC 2735  &   1 & $-21.00$ & A,TT,I,C  \\
NGC 1255  &   6 & $-20.50$ & A   &          NGC 2748  &   4 & $-20.51$ & A	     \\
NGC 1258  &   2 & $-18.61$ & C   &          NGC 2750  &   4  &  $-20.98$  & A  \\
NGC 1300  &   3 & $-21.08$ & A    &         NGC 2764  &   1  &  $-20.67$  & A	\\ 
NGC 1309  &   4 &  $-20.78$  & E &          NGC 2770  &   5 & $-20.49$  & A,C  \\
NGC 1313  &   7 &  $-18.60$  & A,E &        NGC 2776  &   5 & $-20.66$  & A	\\	
\hline
\end{tabular} 
\end{table*}

\setcounter{table}{0}
\begin{table*}
\caption{Continued.}
\begin{tabular}{lccrlccr}
\hline
Name & T-Type & Abs.3.6 $\mu$m & Classification & Name & T-Type & Abs. 3.6 $\mu$m & Classification \\
 &  & AB Mag. &  &  &  & AB Mag. &  \\
\hline
NGC 2782  &   1 & $-20.61$  & E,S &	   NGC 3424  &   2 & $-20.91$  & A,C 	  \\ 	 
NGC 2793  &  99 & $-18.49$  & A &	   NGC 3430  &   4 & $-20.99$  & A,C 	  \\ 	
NGC 2798  &   1 & $-20.68$  & A,I,C &	   NGC 3432  &   9 & $-19.03$  & A,C 	  \\ 	
NGC 2799  &   8 & $-18.51$  & A,I,C &	   NGC 3433  &   3 & $-20.93$  & TT? 	  \\ 	
NGC 2805  &   5 & $-20.44$  & E &	   NGC 3440  &  10 & $-19.11$  & A 	  \\ 	
NGC 2814  &  10 & $-18.65$  & A,C &	   NGC 3443  &   8 & $-17.72$  & E 	  \\ 	
NGC 2820  &   6 & $-20.43$  & C &	   NGC 3445  &   9 & $-18.89$  & C 	  \\ 	
NGC 2854  &   2 & $-20.14$  & A,C &	   NGC 3447A &   9 & $-18.12$  & A,TT?,C  \\ 	
NGC 2894  &$-2$ & $-21.35$  & U &	   NGC 3447B &  10 & $-15.20$  & A,C 	  \\ 	
NGC 2964  &   3 & $-20.84$  & C &	   NGC 3448  &  10 & $-20.06$  & A,C 	  \\ 	
NGC 2966  &   1 & $-19.99$  & A,C &	   NGC 3455  &   5 & $-19.45$  & A 	  \\ 	
NGC 2968  &$-1$ & $-19.78$  & C &	   NGC 3468  &  99 & $-22.72$  & A 	  \\ 	
NGC 2986  &$-5$ & $-22.54$  & C &	   NGC 3471  &   0 & $-20.02$  & C 	  \\ 	
NGC 3003  &   7 & $-19.91$  & A,TT?&	   NGC 3485  &   4 & $-19.52$  & A 	  \\ 	
NGC 3018  &   8 & $-18.37$  & A,C &	   NGC 3488  &   6 & $-20.39$  & U 	  \\ 	
NGC 3020  &   9 & $-19.35$  & A,TT? &	   NGC 3495  &   5 & $-20.21$  & A 	  \\ 	
NGC 3023  &   8 & $-19.50$  & A,C &	   NGC 3510  &   7 & $-17.42$  & A,W 	  \\	   	 
NGC 3024  &   8 & $-18.02$  & A &	   NGC 3513  &   6 & $-18.93$  & A 	  \\	   	 
NGC 3026  &   7 & $-18.82$  & A &	   NGC 3521  &   4 & $-22.21$  & A 	  \\	   	 
NGC 3027  &   8 & $-18.90$  & A,TT? &	   NGC 3526  &   8 & $-19.46$  & U 	  \\	   	 
NGC 3034  &   0 & $-21.12$  & A &	   NGC 3547  &   6 & $-19.29$  & A 	  \\	   	 
NGC 3044  &   8 & $-20.48$  & A &	   NGC 3583  &   2 & $-21.48$  & C 	  \\ 	   	 
NGC 3049  &   2 & $-20.01$  & A &	   NGC 3589  &  10 & $-18.04$  & A 	  \\	   	 
NGC 3057  &   8 & $-18.51$  & A &	   NGC 3596  &   4 & $-19.78$  & A,TT?     \\	   	 
NGC 3061  &   3 & $-20.56$  & A &	   NGC 3600  &  1 & $-18.29$  & A  \\		       
NGC 3065  &$-2$ & $-20.96$  & C &	   NGC 3619  &$-5$& $-21.07$  & S    \\ 	   
NGC 3066  &   2 & $-20.03$ & C   &         NGC 3625  &  7 & $-19.39$  & A    \\ 	   
NGC 3073  &$-3$ & $-18.63$ & C   &         NGC 3627  &  3 & $-21.70$  & A,TT  \\	   
NGC 3079  &  3 & $-21.82$  & A,C,E,W? &	   NGC 3628  &  4 & $-21.73$  & A,W   \\		       
NGC 3094  &  1 & $-21.77$  & A?  &	   NGC 3631  &  5 & $-20.14$  & E    \\ 	      
NGC 3104  & 10 & $-16.48$  & A   &	   NGC 3633  &  1 & $-20.87$  & A     \\	      
NGC 3118  &  6 & $-18.33$  & A   &         NGC 3642  &  2 & $-20.71$  & E      \\	     
NGC 3153  &  7 & $-19.90$  & A   &         NGC 3652  &  5 & $-19.94$  & A      \\	     
NGC 3162  &  5 & $-20.35$  & A   &         NGC 3664  &  9 & $-18.59$  & A,TT,M?  \\	     
NGC 3165  &  9 & $-17.18$  & C   &         NGC 3666  &  4 & $-19.76$  & W?	  \\	     
NGC 3169  &  2 & $-21.70$  & A   &         NGC 3669  &  7 & $-20.63$  & A	\\	     
NGC 3185  &  1 & $-20.19$  & A,C &         NGC 3672  &  5 & $-21.49$  & A	 \\	   
NGC 3187  &  5 & $-19.53$  & A   &    	   NGC 3675  &  3 & $-21.84$  & E	 \\	   
NGC 3190  &  1 & $-21.80$  & A,C &	   NGC 3686  &  4 & $-20.33$  & A	 \\	   
NGC 3206  &  7 & $-18.69$  & A   &    	   NGC 3687  &  2 & $-19.84$  & C      \\	   
NGC 3227  &  1 & $-21.55$  & A,I,C   &	   NGC 3701  &  5 & $-20.23$  & A      \\	   
NGC 3239  &  9 & $-17.36$  & A,TT,M  &     NGC 3705  &  3 & $-20.84$  & A     \\	   
NGC 3245A &  7 & $-17.75$  & A &      	   NGC 3712  & 10 & $-17.00$  & A     \\	   
NGC 3246  &  7 & $-19.37$  & A &      	   NGC 3718  &  1 & $-20.88$  & A,TT \\     	   
NGC 3264  &  8 & $-17.72$  & A &      	   NGC 3726  &  4 & $-20.87$  & A     \\	   
NGC 3294  &  4 & $-21.54$  & A &      	   NGC 3729  &  0 & $-20.41$  & A     \\		   
NGC 3306  &  3 & $-20.75$  & TT &     	   NGC 3733  &  5 & $-19.27$  & TT,U \\         	   
NGC 3310  &  4 & $-20.59$  & A,E,S &	   NGC 3735  &  5 & $-22.04$  & A     \\	
NGC 3320  &  5 & $-20.31$  & U &      	   NGC 3755  &  5 & $-19.74$  & A     \\		   
NGC 3321  &  6 & $-20.23$  & A &      	   NGC 3769  &  6 & $-19.38$  & C     \\		   
NGC 3338  &  4 & $-21.10$  & A &      	   NGC 3779  &  9 & $-17.18$  & A     \\
NGC 3359  &  7 & $-20.41$  & A &      	   NGC 3780  &  5 & $-21.77$  & A     \\	       
NGC 3364  &  4 & $-20.40$  & A &      	   NGC 3782  &  9 & $-17.68$  & U     \\	       
NGC 3365  &  7 & $-18.65$  & A &      	   NGC 3786  &  0 & $-21.50$  & I,C  \\ 	       
NGC 3368  & $-1$ & $-21.35$  & TT,U &      NGC 3788  &   1 & $-21.67$  & A,I,C \\    
NGC 3377A & 10 & $-15.30$  & C & 	   NGC 3846A &   9 & $-18.43$  & A,TT \\	  
NGC 3381  &   8 & $-19.45$  & A    &	   NGC 3850  &   9 & $-17.61$  & A   \\ 	
NGC 3384  &$-3$ & $-20.74$  & TT,C &	   NGC 3876  &   8 & $-19.68$  & C  \\  	
NGC 3389  &   5 & $-19.77$  & A   &	   NGC 3877  &   4 & $-20.72$  & A   \\    
NGC 3395  &   5 & $-19.90$  & A,TT?,I,C &  NGC 3885  &  $-1$ & $-21.29$  & U  \\   		   
NGC 3396  &  10 & $-19.61$  & A,I,C &	   NGC 3887  &   4 & $-20.87$  & A  \\	  		   
NGC 3414  &$-3$ & $-21.75$  & C    &	   NGC 3888  &   3 & $-21.27$  & A  \\	  		
\hline
\end{tabular} 
\end{table*}

\setcounter{table}{0}
\begin{table*}
\caption{Continued.}
\begin{tabular}{lccrlccr}
\hline
Name & T-Type & Abs.3.6 $\mu$m & Classification & Name & T-Type & Abs. 3.6 $\mu$m & Classification \\
 &  & AB Mag. &  &  &  & AB Mag. &  \\
\hline
NGC 3896  &   9 & $-17.44$  & C,U &	           NGC 4283  &  $-5$ & $-19.30$  & C	      \\ 
NGC 3912  &   9 & $-19.71$  & A &	   	   NGC 4288  &   9 & $-17.75$  & A	      \\
NGC 3917  &   5 & $-19.62$  & C &	   	   NGC 4293  &   0 & $-20.72$  & A	      \\
NGC 3930  &   7 & $-18.64$  & E &	   	   NGC 4294  &   7 & $-19.19$  & C	      \\
NGC 3938  &   5 & $-20.98$  & A &	   	   NGC 4298  &   4 & $-20.17$  & I?,C	      \\
NGC 3949  &   6 & $-20.36$  & A? &	   	   NGC 4299  &   9 & $-18.36$  & C	      \\
NGC 3952  &  10 & $-18.49$  & A &	   	   NGC 4302  &   4 & $-21.41$  & I?,E,C,W?    \\
NGC 3956  &   6 & $-19.85$  & A,C &	   	   NGC 4303A &   8 & $-17.68$  & E	      \\
NGC 3962  &  $-4$ & $-22.35$  & U &	   	   NGC 4309  &  $-1$ & $-18.46$  & C	      \\
NGC 3972  &   5 & $-19.39$  & E &	   	   NGC 4313  &   1 & $-19.94$  & C	      \\
NGC 3976  &   4 & $-21.58$  & A &	   	   NGC 4319  &   1 & $-20.83$  & TT,C	      \\ 
NGC 3981  &   4 & $-20.97$  & A &	   	   NGC 4321  &   4 & $-21.99$  & C	      \\
NGC 3982  &   3 & $-20.40$  & A &	   	   NGC 4343  &   1 & $-20.65$  & C	      \\
NGC 3992  &   2 & $-22.39$  & C &	   	   NGC 4355  &   0 & $-19.31$  & C	      \\
NGC 3998  &  $-2$ & $-21.70$  & C &	   	   NGC 4388  &   2 & $-21.26$  & C	      \\ 
NGC 4010  &   8 & $-19.54$  & A &	   	   NGC 4393  &   9 & $-16.99$  & U	      \\
NGC 4020  &   8 & $-16.73$  & A &	   	   NGC 4394  &   0 & $-20.53$  & TT,C	      \\
NGC 4027  &   8 & $-21.25$  & A,C &	   	   NGC 4395  &   8 & $-17.04$  & A	      \\
NGC 4038  &  99 & $-21.97$  & TT,I,C &	   	   NGC 4402  &   5 & $-20.28$  & C	      \\
NGC 4039  &  99 & $-20.79$  & TT,I,C &     	   NGC 4406  &  $-4$ & $-999  $  & U	      \\
NGC 4049  &   9 & $-16.78$  & A &	   	   NGC 4411A &   7 & $-17.75$  & TT,C	      \\
NGC 4051  &   3 & $-20.84$  & A &	   	   NGC 4423  &   9 & $-18.03$  & W	      \\
NGC 4088  &  5 & $-21.22$  & A  &          	   NGC 4438  &   0 & $-21.22$  & A,C	    \\
NGC 4094  &  6 & $-19.87$  & U  &          	   NGC 4472  &$-5$ & $-23.12$  & C	    \\
NGC 4096  &  7 & $-20.24$  & A  &          	   NGC 4485  &  10 & $-17.47$  & A,I,C     \\
NGC 4105  &$-5$& $-999  $  & I,U,C  &          	   NGC 4488  &   1 & $-18.92$  & A,TT?     \\
NGC 4106  &  1 & $-22.40$  & A,I,C &	   	   NGC 4490  &   9 & $-19.73$  & A,I,C       \\   
NGC 4108  &  5 & $-20.67$  & C,U &	   	   NGC 4496A &   7 & $-19.19$  & U	      \\  
NGC 4111  &$-3$ & $-20.70$  & E     &	   	   NGC 4503  &  10 & $-20.54$  & C	      \\   
NGC 4117  &$-3$ & $-18.81$  & C     &	   	   NGC 4517A &   8 & $-18.71$  & A	      \\   
NGC 4123  &   3 & $-20.41$  & A     &	   	   NGC 4519  &   7 & $-20.21$  & A	      \\   
NGC 4141  &   8 & $-17.92$  & TT    &	           NGC 4523  &   9 & $-17.39$  & U	      \\     
NGC 4144  &   9 & $-17.25$  & W     &	   	   NGC 4532  &  10 & $-19.06$  & A	      \\   
NGC 4151  &   0 & $-19.87$  & E?    &	   	   NGC 4533  &   7 & $-18.38$  & C	      \\   
NGC 4157  &   5 & $-21.60$  & A?,E  &	   	   NGC 4534  &   9 & $-17.71$  & A	      \\   
NGC 4163  &  11 & $-13.83$  & E?    &	   	   NGC 4535  &   5 & $-21.44$  & A	      \\   
NGC 4165  &   3 & $-19.82$  & C?    &	   	   NGC 4536  &   4 & $-21.02$  & A	      \\   
NGC 4173  &   9 & $-15.64$  & A     &	   	   NGC 4559  &   6 & $-19.69$  & A	      \\   
NGC 4183  &   6 & $-18.86$  & E,W   &	   	   NGC 4561  &   7 & $-17.36$  & A	      \\   
NGC 4190  &  10 & $-14.09$  & E?    &	   	   NGC 4562  &   8 & $-16.82$  & A?	      \\   
NGC 4192  &   2 & $-21.54$  & A     &	   	   NGC 4567  &   4 & $-21.05$  & I?,C	      \\   
NGC 4193  &   2 & $-20.89$  & A?    &	   	   NGC 4568  &   5 & $-21.57$  & I?,C	      \\   
NGC 4194  &   1 & $-21.45$  & E     &	   	   NGC 4571  &   5 & $-20.29$  & C,U	      \\   
NGC 4197  &   8 & $-19.79$  & E     &	   	   NGC 4572  &   6 & $-19.16$  & A	      \\   
NGC 4204  &   8 & $-16.67$  & A     &	   	   NGC 4594  &$-1$ & $-22.64$  & U	      \\   
NGC 4212  &   3 & $-20.63$  & A     &	   	   NGC 4597  &   8 & $-18.65$  & A,C	      \\   
NGC 4216  &   2 & $-22.03$  & C     &	   	   NGC 4605  &  10 & $-18.27$  & A,U	      \\   
NGC 4217  &   5 & $-21.45$  & U     &	   	   NGC 4625  &   9 & $-17.28$  & A	      \\   
NGC 4222  &   7 & $-19.43$  & C     &	   	   NGC 4631  &   7 & $-20.20$  & A,E?,C      \\    
NGC 4224  &$-1$ & $-21.85$  & C     &	   	   NGC 4633  &   6 & $-18.46$  & C	      \\   
NGC 4234  &   9 & $-19.87$  & A     &	   	   NGC 4634  &   7 & $-16.42$  & A,C	      \\   
NGC 4235  &$-1$ & $-20.98$  & A     &	   	   NGC 4636  &  $-4$ & $-999  $  & U	      \\   
NGC 4236  &   9 & $-999  $& A,E    &	   	   NGC 4639  &   2 & $-20.42$  & C	      \\    
NGC 4237  &   3 & $-20.54$  & E    &	   	   NGC 4647  &   6 & $-999  $  & A,C,U        \\
NGC 4238  &   5 & $-19.34$  & C    &	   	   NGC 4651  &   3 & $-21.73$  & TT	  \\
NGC 4244  &   7 & $-17.85$  & U    &	   	   NGC 4653  &   5 & $-20.58$  & A?,C	  \\
NGC 4252  &   7 & $-16.14$  & A    &	   	   NGC 4654  &   6 & $-20.91$  & A	  \\ 	      
NGC 4254  &   5 & $-21.61$  & A    &       	   NGC 4656  &   8 & $-17.16$ & A		 \\
NGC 4256  &   0 & $-22.05$  & C    &               NGC 4659  &$-2$ & $-17.47$ & U    \\  
NGC 4258  &   2 & $-21.33$  & U    &       	   NGC 4666  &   5 & $-21.90$ & C  \\
NGC 4268  &  $-1$ & $-20.13$  & C  &       	   NGC 4688  &   8 & $-18.37$ & A,E? \\
NGC 4273  &   5 & $-21.41$  & A,C  &       	   NGC 4698  &   0 & $-21.69$ & E? \\
\hline
\end{tabular} 
\end{table*}

\setcounter{table}{0}
\begin{table*}
\caption{Continued.}
\begin{tabular}{lccrlccr}
\hline
Name & T-Type & Abs.3.6 $\mu$m & Classification & Name & T-Type & Abs. 3.6 $\mu$m & Classification \\
 &  & AB Mag. &  &  &  & AB Mag. &  \\
\hline
NGC 4700  &   9 & $-18.69$ & A   &	                 NGC 5477  &  10 & $-14.64$  & U       \\
NGC 4707  &  10 & $-14.78$ & U   &	        	 NGC 5480  &   7 & $-20.46$  & A,C     \\  
NGC 4723  &  10 & $-16.30$ & U &	        	 NGC 5481  &$-4$ & $-20.50$  & C       \\ 
NGC 4725  &   1 & $-21.76$ & A &	        	 NGC 5506  &   0 & $-21.18$  & C,U     \\  
NGC 4731  &   7 & $-19.85$ & A &	                 NGC 5529  &   2 & $-22.20$  & W       \\  	       
NGC 4747  &   9 & $-18.36$ & A,M &	     		 NGC 5534  &   1 & $-20.37$  & C       \\
NGC 4762  &$-2$ & $-21.64$ & U &	     		 NGC 5560  &   7 & $-19.65$  & A,C     \\
NGC 4775  &   6 & $-20.29$ & U &	     		 NGC 5566  &   1 & $-21.79$  & A,C     \\
NGC 4789A &  10 & $-12.69$ & A &	     		 NGC 5569  &   9 & $-17.45$  & E,C     \\
NGC 4793  &   5 & $-21.41$ & A,C &	     		 NGC 5574  &   0 & $-19.86$  & I,C    \\ 
NGC 4795  &   1 & $-21.40$ & A,TT?,I?,C&     		 NGC 5576  &$-5$ & $-21.49$  & I,C   \\ 
NGC 4802  &$-2$ & $-18.86$ & U &	     		 NGC 5584  &   7 & $-20.31$  & A       \\
NGC 4809  &  10 & $-17.26$ & A &	     		 NGC 5597  &   7 & $-21.23$  & A     \\ 
NGC 4814  &   4 & $-21.49$ & TT? &	     		 NGC 5600  &   8 & $-22.92$  & A?     \\
NGC 4899  &   5 & $-20.72$ & A &	     		 NGC 5608  &  10 & $-17.27$  & A?     \\
NGC 4902  &   3 & $-22.21$ & A &	     		 NGC 5636  &   0 & $-18.87$  & C      \\
NGC 4904  &   6 & $-19.91$ & A &	     		 NGC 5645  &   8 & $-19.43$  & A      \\
NGC 4948A &   9 & $-18.30$ & E? &	     		 NGC 5660  &   5 & $-21.07$  & E      \\
NGC 4951  &   3 & $-19.62$  & U     &	     		 NGC 5661  &   6 & $-19.93$  & A,C      \\
NGC 4958  &$-1$ & $-21.13$  & A     &	     		 NGC 5665  &   7 & $-20.41$  & A?     \\
NGC 4961  &   5 & $-20.04$  & C     &	     		 NGC 5669  &   7 & $-20.01$  & A      \\
NGC 4981  &   4 & $-20.82$  & TT    &	     		 NGC 5678  &   3 & $-21.61$  & E,C    \\
NGC 4995  &   2 & $-21.50$  & U    &	     		 NGC 5708  &   5 & $-20.05$  & A     \\ 
NGC 5018  &$-4$ & $-22.62$  & E,C  &	     		 NGC 5719  &   0 & $-21.35$  & A    \\ 
NGC 5022  &   3 & $-21.10$  & A,C   &	     		 NGC 5730  &   7 & $-19.64$  & W?     \\
NGC 5033  &   6 & $-22.04$  & A?    &	     		 NGC 5731  &   9 & $-18.94$  & A    \\ 
NGC 5042  &   6 & $-19.52$  & U     &	     		 NGC 5740  &   2 & $-20.81$  & TT      \\
NGC 5054  &   4 & $-21.38$  & C    &	     		 NGC 5746  &   0 & $-22.81$  & U       \\
NGC 5055  &   4 & $-21.59$  & E,U  &	     		 NGC 5750  &   0 & $-21.44$  & A       \\
NGC 5078  &   3 & $-22.71$  & C     &	     		 NGC 5757  &   2 & $-21.59$  & C       \\
NGC 5079  &   4 & $-20.05$  & C           &  		 NGC 5768  &   5 & $-19.99$  & A       \\
NGC 5084  &   0 & $-21.94$  & A       	  &  		 NGC 5775  &   5 & $-21.60$  & C       \\
NGC 5085  &   4 & $-21.73$  & A       	  &  		 NGC 5777  &   0 & $-21.75$  & C       \\
NGC 5103  &$-3$ & $-19.51$  & U       	  &  		 NGC 5792  &   2 & $-21.60$  & U       \\
NGC 5107  &  10 & $-17.72$  & A       	 &   		 NGC 5809  &   1 & $-20.23$  & A       \\
NGC 5112  &   7 & $-20.08$  & A       	 &   		 NGC 5846  &$-4$ & $-22.69$  & C,U    \\ 
NGC 5122  &$-3$ & $-20.28$  & PR?     	  &  		 NGC 5850  &   2 & $-21.32$  & C     \\ 
NGC 5145  &$-1$ & $-21.26$  & U       	  &  		 NGC 5892  &   6 & $-20.48$  & A?      \\
NGC 5169  &   5 & $-19.40$  & C       	  &  		 NGC 5900  &   3 & $-21.95$  & TT    \\ 
NGC 5194  &   4 & $-21.93$  & I,C     	 &   		 NGC 5915  &   8 & $-20.36$  & C	\\
NGC 5195  &   0 & $-20.57$  & A,E,I,C 	 &   		 NGC 5916  &   1 & $-20.53$  & A	\\
NGC 5205  &   2 & $-19.75$  & E?      	  &  		 NGC 5916A &  10 & $-18.60$  & A	\\
NGC 5216  &$-5$ & $-21.25$  & E?,I,C     &   		 NGC 5921  &   3 & $-20.72$  & A?,U	\\
NGC 5218  &   1 & $-21.86$  & S,I,C     &    		 NGC 5930  &   0 & $-20.96$  & C,I	\\
NGC 5240  &   3 & $-20.69$  & U        &     		 NGC 5953  &$-1$ & $-20.62$  & I,C	\\
NGC 5247  &   5 & $-21.81$  & E        &     		 NGC 5954  &  3  & $-20.19$  & I,C	\\
NGC 5248  &   4 & $-21.43$  & U         &    		 NGC 5963  &   5 & $-19.89$  & A?,U  \\
NGC 5297  &   4 & $-21.18$  & C        &     		 NGC 5981  &$-1$ & $-20.48$  & C?   \\
NGC 5320  &   5 & $-20.51$  & A     &	     		 NGC 5985  &   3 & $-22.36$  & TT? \\	 
NGC 5334  &   6 & $-20.68$  & TT?   &	     		 NGC 6012  &   2 & $-19.83$  & U   \\
NGC 5348  &   7 & $-18.36$  & W?    &	     		 NGC 6070  &	 5 & $-21.55$  & A   \\
NGC 5350  &   3 & $-21.16$  & C,U   &	     		 NGC 6140  & 7 & $-19.32$  & E,M?   \\
NGC 5353  &$-1$ & $-22.35$  & I,C  &	     	    	 NGC 6168  & 8 & $-19.98$  & A   \\
NGC 5354  &$-4$ & $-22.16$  & I,C  &	     		 NGC 6237  &   9 & $-17.58$  & E   \\ 
NGC 5355  &$-3$ & $-19.90$  & C     &	     		 NGC 6239  & 9 & $-19.32$  & A   \\
NGC 5383  &   2 & $-21.71$  & A,E,C &	     		 NGC 6278  &$-2$ & $-21.21$  & C   \\ 
NGC 5403  &   2 & $-21.63$  & W,C   &	     		 NGC 6340  &  0 & $-20.92$  & C,U   \\
NGC 5426  &   5 & $-21.28$  & I,C  &	     		 NGC 6861E &  8 & $-18.59$  & A     \\
NGC 5427  &   4 & $-21.60$  & I,C  &    	  	 NGC 6925  &  3 & $-21.82$  & E?     \\	  
NGC 5448  &   1 & $-21.50$  & A    &   	  	         NGC 7059  &  7 & $-20.61$  & A      \\  
NGC 5457  &   5 & $-21.38$  & TT   &   	  	         NGC 7064  &  8 & $-17.10$  & A     \\	 
NGC 5468  &   6 & $-21.29$  & E    &   	  	         NGC 7090  &  8 & $-18.93$  & U    \\	 
\hline                                                  
\end{tabular} 			                        
\end{table*}			                        
				                        
\setcounter{table}{0}
\begin{table*}
\caption{Continued.}
\begin{tabular}{lccrlccr}
\hline
Name & T-Type & Abs.3.6 $\mu$m & Classification & Name & T-Type & Abs. 3.6 $\mu$m & Classification \\
 &  & AB Mag. &  &  &  & AB Mag. &  \\
\hline 
NGC 7140  &  3 & $-21.40$  & E     &	        	UGC 4169  &  6 & $-19.08$ & E	     \\ 	 
NGC 7162A &  8 & $-19.11$  & U      &	        	UGC 4238  &  8 & $-18.38$ & U	     \\ 	 
NGC 7167  &  6 & $-19.78$  & U     &	        	UGC 4305  & 10 & $-15.58$ & A	     \\ 	 
NGC 7183  &  0 & $-21.79$  & A    &	        	UGC 4393  & 10 & $-18.04$ & A	     \\ 	 
NGC 7188  &  3 & $-18.59$  & A &	          	UGC 4426  & 10 & $-14.07$ & A	    \\ 
NGC 7213  & $-2$ & $-22.25$  & C &	          	UGC 4483  & 10 & $-12.05$ & A	   \\  
NGC 7307  &  7 & $-19.62$  & A &	          	UGC 4499  &  9 & $-16.62$ & E	   \\  
NGC 7347  &  6 & $-19.96$  & W? &	          	UGC 4543  &  9 & $-17.91$ & A	  \\   
NGC 7361  &  8 & $-18.69$  & A &	          	UGC 4551  & $-1$ & $-20.14$ & A   \\   
NGC 7412  &  4 & $-19.24$  & A &	          	UGC 4704  &  8 & $-15.38$ & W	  \\   
NGC 7418A &  3 & $-17.82$  & E &	          	UGC 4722  &  9 & $-16.60$ & A,TT  \\   
NGC 7424  &  6 & $-19.60$  & A?,U &	          	UGC 4797  & 10 & $-17.19$ & U	  \\   
NGC 7456  &  6 & $-19.12$  & A &	          	UGC 4834  &  10 & $-16.74$ & A     \\		
NGC 7462  &  9 & $-18.42$  & U  &                 	UGC 4837  &  10 & $-17.26$ & U     \\		
NGC 7463  &  6  & $-20.13$  & E,C    &	          	UGC 4841  &   6 & $-19.05$ & U     \\		
NGC 7465  &$-2$ & $-20.17$  & A,C    &	          	UGC 4867  &   7 & $-18.36$ & U     \\		
NGC 7479  &  3  & $-22.31$  & E      &	          	UGC 4871  &   9 & $-17.27$ & U     \\		
NGC 7496  &  3  & $-19.73$  & A?,U   &	          	UGC 4970  &   7 & $-19.22$ & W   \\	
NGC 7531  &  1  & $-20.73$  & TT    &	          	UGC 5050  &   8 & $-18.71$ & TT      \\ 	
NGC 7537  &  5  & $-20.33$  & C     &	          	UGC 5139  &  10 & $-999  $ & U       \\
NGC 7541  &  5  & $-21.96$  & C      &	          	UGC 5179  &  $-2$ & $-17.14$ & C     \\
NGC 7552  &  1  & $-21.33$  & A      &	          	UGC 5336  &  10 & $-12.66$ & U       \\
NGC 7590  &  4  & $-21.13$  & E      &	          	UGC 5340  &  10 & $-13.29$ & A       \\
NGC 7599  &  6  & $-20.61$  & A,C   &	          	UGC 5364  &  10 & $-999  $ & U       \\
NGC 7625  &  1  & $-20.44$  & A            &      	UGC 5391  &   8 & $-17.19$ & A,U       \\	
NGC 7694  & 10  & $-19.19$  & C 	   &      	UGC 5421  &  10 & $-14.71$ & U         \\  	   
NGC 7714  &  1  & $-20.62$  & TT,C	   &      	UGC 5459  &   8 & $-19.49$ & E         \\ 
NGC 7715  & 99  & $-18.34$  & E,C	   &      	UGC 5464  &  10 & $-15.35$ & U         \\
NGC 7727  &  1  & $-21.66$  & TT?,S?,M?,U  &       	UGC 5478  &   9 & $-17.31$ & U         \\ 
NGC 7731  &  2  & $-19.56$  & C 	   &          	UGC 5522  &   5 & $-18.78$ & A         \\ 
NGC 7732  &  8  & $-19.37$  & A,C  & 	              	UGC 5571  &  10 & $-13.84$ & U         \\ 
NGC 7741  &  6  & $-19.28$  & U    & 	          	UGC 5633  &   7 & $-17.73$ & U         \\ 
NGC 7755  &  4  & $-21.17$  & A    & 	          	UGC 5677  &  10 & $-15.85$ & E?        \\ 
NGC 7757  &  6  & $-20.34$  & E    & 	          	UGC 5688  &  10 & $-18.53$ & A,E?     \\ 
NGC 7800  & 10  & $-18.79$  & A    & 	          	UGC 5689  &   7 & $-20.01$ & W?      \\ 
UGC 17    & 10  & $-15.69$ & U     & 	          	UGC 5707  &   7 & $-18.19$ & A         \\ 
UGC 99    &  9  & $-17.33$ & U    & 	          	UGC 5708  &   8 & $-17.21$ & E       \\ 
UGC 156   & 10  & $-16.83$ & A    & 	          	UGC 5764  &  10 & $-13.00$ & E       \\
UGC 191   &  8  & $-16.81$ & A?   & 	          	UGC 5791  &   9 & $-15.67$ & E        \\
UGC 260   &  6  & $-19.78$ & A,C  & 	          	UGC 5829  &  10 & $-15.64$ & E  	 \\
UGC 634   & 10  & $-17.28$ & U    & 	          	UGC 5832  &  9  & $-17.62$ & A  	 \\
UGC 711   &  9  & $-17.63$ & U    & 	          	UGC 5844  &  9 & $-16.07$ & A	  \\
UGC 882   & 10  & $-17.40$ & A,U  &               	UGC 5918  & 10 & $-13.55$ & U	  \\
UGC 891   &  9 & $-14.94$ & A  &	          	UGC 5947  & 10 & $-16.65$ & A	  \\
UGC 903   &  2 & $-21.33$ & A  &                  	UGC 5958  &  7 & $-18.20$ & C	  \\
UGC 941   & 10 & $-17.47$ & A?,U &	          	UGC 5979  & 10 & $-16.54$ & E	  \\
UGC 958   &  8 & $-17.15$ & W	&	          	UGC 5989  &  8 & $-17.08$ & A	  \\
UGC 1014  & 10 & $-17.53$ & I	&	          	UGC 6104  &  6 & $-18.97$ & TT?   \\
UGC 1133  & 10 & $-16.14$ & U	&	          	UGC 6145  & 10 & $-13.73$ & U	 \\ 
UGC 1176  & 10 & $-15.39$ & A	&	          	UGC 6151  & 10 & $-16.86$ & U	 \\ 
UGC 1195  &  9 & $-16.53$ & A	      &           	UGC 6157  &  8 & $-18.72$ & A	 \\ 
UGC 1197  &  9 & $-17.80$ & A	    &	          	UGC 6171  &  9 & $-16.87$ & U	 \\ 
UGC 1547  &  9 & $-18.30$ & E?,C,U  &	          	UGC 6181  & 10 & $-15.94$ & U	 \\ 
UGC 1670  &  9 & $-16.40$ & U	    &	          	UGC 6307  &  9 & $-18.26$ & I?  \\  
UGC 1839  &  9 & $-16.61$ & A?       &            	UGC 6309  &  5  & $-20.70$ & A    \\
UGC 1862  &  5 & $-18.21$ & U	    &	          	UGC 6341  & 10  & $-15.97$ & C    \\
UGC 1981  & 10 & $-16.21$ & U	    &	          	UGC 6345  & 10  & $-17.78$ & A,E? \\
UGC 2275  & 10 & $-15.95$ & U	    &	          	UGC 6355  &  8  & $-17.92$ & C   \\  
UGC 2302  &  9 & $-16.44$ & U	    &                   UGC 6378  &  8  & $-17.17$ & E?  \\
UGC 2345  &  9 & $-16.53$ & A	    &  	  		UGC 6433  & 10  & $-17.90$ & A?  \\
UGC 3070  &  9 & $-18.17$ & U	    &  	  		UGC 6446  &  7  & $-16.97$ & U   \\
UGC 4121  &  9 & $-16.18$ & U	    &  	  		UGC 6534  &  9  & $-17.95$ & A \\
\hline
\end{tabular} 
\end{table*}

\setcounter{table}{0}
\begin{table*}
\caption{Continued.}
\begin{tabular}{lccrlccr}
\hline
Name & T-Type & Abs.3.6 $\mu$m & Classification & Name & T-Type & Abs. 3.6 $\mu$m & Classification \\
 &  & AB Mag. &  &  &  & AB Mag. &  \\
\hline
UGC 6628  &  9  & $-17.85$ & U &     		     UGC 8246  &  8  & $-16.98$ & A,E \\      
UGC 6670  &  8  & $-17.63$ & E &     		     UGC 8303  & 10  & $-17.44$ & A  \\ 
UGC 6682  &  9  & $-17.25$ & U &     		     UGC 8320  & 10  & $-14.60$ & U  \\ 
UGC 6747  & 10  & $-15.39$ & W &     		     UGC 8365  & 10  & $-17.18$ & E  \\ 
UGC 6780  &  6  & $-18.64$ & E &     		     UGC 8441  & 10  & $-17.47$ & U  \\ 
UGC 6782  & 10  & $-14.64$ & U &     		     UGC 8449  &  8  & $-16.28$ & A  \\ 
UGC 6816  &  9  & $-17.15$ & A &     		     UGC 8489  & 9   & $-17.00$ & TT? \\
UGC 6817  & 10  & $-12.81$ & U &     		     UGC 8508  & 10  & $-12.50$ & E?  \\
UGC 6849  &  8  & $-16.54$ & A  &    		     UGC 8597  &  8  & $-18.60$ & TT?  \\
UGC 6903  &   6 & $-19.46$ & U     & 		     UGC 8614  & 10  & $-18.77$ & U   \\     
UGC 6912  &  10 & $-17.25$ & U     & 		     UGC 8630  &  9  & $-19.16$ & A   \\     
UGC 6955  &  10 & $-17.45$ & U     & 		     UGC 8639  & 10  & $-17.48$ & A   \\     
UGC 6956  &   9 & $-16.38$ & E     & 		     UGC 8642  &  9  & $-16.68$ & A   \\     
UGC 6969  &  10 & $-16.82$ & A,C   & 		     UGC 8688  & 10  & $-17.23$ & U   \\ 
UGC 6983  &   7 & $-18.31$ & A?    & 		     UGC 8726  &  8  & $-17.49$ & E   \\
UGC 7019  &  10 & $-16.96$ & C,U   & 		     UGC 8733  &  8  & $-18.73$ & C   \\
UGC 7053  &  10 & $-16.26$ & U   &   		     UGC 8877  &  8  & $-17.58$ & C,U \\
UGC 7089  &   9 & $-17.21$ & A	   & 		     UGC 8892  &  8  & $-17.58$ & U   \\
UGC 7094  &  10 & $-15.77$ & E    &  		     UGC 8995  &  6  & $-17.73$ & TT? \\
UGC 7125  &  10 & $-16.82$ & E   &   		     UGC 9057  &  7  & $-18.33$ & A,TT \\
UGC 7170  &   7 & $-18.02$ & W?   &  		     UGC 9126  & 10  & $-17.51$ & U   \\
UGC 7175  &  10 & $-16.56$ & A    &  		     UGC 9128  & 10  & $-12.57$ & U   \\
UGC 7189  &   7 & $-18.52$ & U    &  		     UGC 9169  &  9  & $-16.76$ & A,TT? \\
UGC 7218  &  10 & $-16.22$ & U    &  		     UGC 9206  & 10  & $-18.59$ & U   \\
UGC 7242  &  99 & $-999  $ & U    &  		     UGC 9242  &  6  & $-17.11$ & E   \\
UGC 7257  &  10 & $-16.67$ & A,C &   		     UGC 9245  &  8  & $-16.68$ & A   \\
UGC 7271  &   8 & $-14.73$ & A    &  		     UGC 9249  &  8  & $-16.52$ & A   \\
UGC 7300  &  10 & $-14.43$ & U    &  		     UGC 9310  &  8  & $-17.94$ & A   \\
UGC 7321  &   7 & $-17.68$ & A   &   		     UGC 9469  &  9  & $-17.05$ & U   \\
UGC 7332  &  10 & $-16.05$ & U    &  		     UGC 9760  &  8  & $-16.38$ & E,W? \\
UGC 7396  &   8 & $-17.68$ & A?   &  		     UGC 9815  &  8  & $-17.70$ & W   \\
UGC 7408  &  10 & $-15.01$ & U    &  		     UGC 9837  &  6  & $-19.40$ & A?,U \\
UGC 7534  &  10 & $-16.45$ & U    &  		     UGC 9845  &  7  & $-17.23$ & U   \\
UGC 7557  &   8 & $-16.97$ & U    &  		     UGC 9856  &  7  & $-17.84$ & W   \\
UGC 7559  &  10 & $-13.68$ & U    &  		     UGC 9858  &  3  & $-20.58$ & E,C \\
UGC 7590  &   9 & $-18.25$ & U  &    		     UGC 9875  &  9  & $-17.72$ & U   \\
UGC 7599  &  10 & $-13.01$ & U    &  		     UGC 9936  &  9  & $-18.05$ & A   \\
UGC 7605  &  10 & $-13.03$ & U   &   		     UGC 10014 & 10  & $-17.06$ & U    \\
UGC 7608  &  10 & $-14.42$ & U    &  		     UGC 10041 &  8  & $-19.04$ & E    \\
UGC 7639  &  11 & $-15.06$ & A   &   		     UGC 10043 &  1  & $-20.75$ & W,C  \\
UGC 7673  & 10  & $-14.77$ & U    &  		     UGC 10054 &  7  & $-18.13$ & A	\\
UGC 7698  & 10  & $-14.97$ & U    &  		     UGC 10061 & 10  & $-17.17$ & A    \\
UGC 7699  &  7  & $-17.62$ & A    &  		     UGC 10194 &  8  & $-16.09$ & A,E  \\
UGC 7700  &  9  & $-18.82$ & C,U  &  		     UGC 10288 &  5  & $-20.51$ & E? \\
UGC 7719  & 10  & $-14.79$ & A    &  		     UGC 10310 & 10  & $-16.72$ & A,C  \\
UGC 7730  &  7  & $-17.51$ & A    &  		     UGC 10437 &  9  & $-18.33$ & E    \\
UGC 7795  & 10  & $-10.74$ & U    &  		     UGC 10445 &  8  & $-18.74$ & A    \\
UGC 7802  &  7  & $-17.59$ & W?   &  		     UGC 10477 &  8  & $-15.54$ & A,W?  \\
UGC 7906  & 10  & $-14.86$ & U    &  		     UGC 10608 & 10  & $-15.57$ & A    \\
UGC 7911  &  8  & $-18.45$ & U    &  		     UGC 10650 & 10  & $-18.33$ & A,U  \\
UGC 7949  & 10  & $-999  $ & U    &  		     UGC 10736 &  8  & $-15.59$ & U    \\
UGC 8040  &  3  & $-19.72$ & A,C  &  		     UGC 10806 &  8  & $-17.59$ & A    \\
UGC 8052  &  5  & $-18.54$ & A?   &  		     UGC 10854 &  9  & $-17.71$ & A    \\
UGC 8053  &  9  & $-16.72$ & U    &  		     UGC 11782 & 10  & $-16.92$ & A    \\
UGC 8056  &  7  & $-17.87$ & U    &  		     UGC 12178 &  8  & $-19.34$ & E	\\
UGC 8084  &  9  & $-18.64$ & U    &  		     UGC 12313 &  9  & $-16.93$ & A?,C \\
UGC 8127  &  9  & $-16.10$ & A,C  &  		     UGC 12350 &  8  & $-17.98$ & U    \\
UGC 8146  &  8  & $-16.94$ & A    &  		     UGC 12578 & 10  & $-17.87$ & A,E \\ 
UGC 8153  &  7  & $-18.54$ & U    &  		     UGC 12613 & 10  & $-13.38$ & U   \\ 
UGC 8155  &  1  & $-19.95$ & E    &  		     UGC 12681 &  9  & $-17.53$ & A   \\
UGC 8166  &  8  & $-17.21$ & E    &  		     UGC 12682 & 10  & $-17.32$ & A   \\ 
UGC 8201  & 10  & $-14.41$ & U    & 	             UGC 12709 &  9  & $-18.41$ & U    \\
\hline
\end{tabular} 
\end{table*}

\setcounter{table}{0}
\begin{table*}
\caption{Continued.}
\begin{tabular}{lccrlccr}
\hline
Name & T-Type & Abs.3.6 $\mu$m & Classification & Name & T-Type & Abs. 3.6 $\mu$m & Classification \\
 &  & AB Mag. &  &  &  & AB Mag. &  \\
\hline
UGC 12732 &  8  & $-16.61$ & U    &                     IC 4901   &   5 & $-20.40$ & A     \\		      
UGC 12843 &  7  & $-18.03$ & U &        	  	IC 4986   &   8 & $-18.17$ & U   \\
UGC 12856 &  9  & $-17.55$ & A &        	  	IC 5007   &   8 & $-20.22$ & A  	 \\
UGC 12857 &  3  & $-19.74$ & W &        	  	IC 5152     & 11 & $-15.31$ & U \\		
IC 167    &  6  & $-18.66$ & TT,C &     	  	IC 5176     &  4 & $-20.90$ & E   \\
IC 223    & 10  & $-16.92$ & A    &     	  	IC 5201     &  7 & $-19.06$ & A   \\
IC 600    &  9  & $-17.89$ & E    &     	  	IC 5249     &  7 & $-18.31$ & A,W	  \\
IC 718    &  9  & $-18.17$ & A     &    	  	IC 5269A    &  9 & $-18.28$ & A   \\
IC 749    &  6  & $-20.51$ & A,C   &      	  	IC 5269C    &  8 & $-17.82$ & A   \\
IC 750    &  1  & $-21.35$ & C     &    	  	IC 5273     &  6 & $-19.77$ & A   \\
IC 755    &  9  & $-18.38$ & A    &     	  	IC 5332     &  6 & $-18.93$ & U   \\
IC 764    &  5  & $-20.04$ & A     &    	  	ESO 011-005 &  8 & $-18.98$ & E        \\
IC 1024   &  9  & $-19.15$ & A     &    	  	ESO 012-010 &  7 & $-18.70$ & A        \\
IC 1029   &  1  & $-21.79$ & C     &    	  	ESO 012-014 & 10 & $-16.89$ & A        \\
IC 1066   &  4  & $-19.52$ & C    &     	  	ESO 015-001 & 10 & $-16.74$ & A        \\
IC 1067   &  3  & $-19.38$ & E,C   &    	  	ESO 027-001 &  4 & $-19.80$ & E        \\
IC 1125   &  8  & $-19.55$ & A     &    	  	ESO 027-008 &  4 & $-20.77$ & A        \\
IC 1151   &  7  & $-20.02$ & A    &     	  	ESO 048-017 &  9 & $-17.63$ & A        \\
IC 1210   &  1  & $-20.29$ & A     &    	  	ESO 054-021 &  8 & $-18.64$ & E        \\
IC 1251   & 10  & $-17.30$ & A    &     	  	ESO 079-003 &  1 & $-21.35$ & A,W?,U  \\
IC 1553   &  6  & $-19.65$  & A    &    	  	ESO 085-047 &  9 & $-16.00$ & U        \\
IC 1555     &  8 & $-17.38$ & A    &    	  	ESO 107-016 &  8 & $-16.95$ & A        \\
IC 1558     &  8 & $-17.34$ & A   &     	  	ESO 114-007 & 10 & $-17.41$ & A        \\
IC 1596     &  4 & $-19.06$ & A    &    	  	ESO 115-021 &  9 & $-15.25$ & A        \\
IC 1613     & 10 & $-999  $ & U    &    	  	ESO 116-012 &  9 & $-999  $ & U        \\
IC 1727     &  9 & $-16.87$ & A,C  &    	  	ESO 119-016 &  8 & $-17.46$ & I?       \\
IC 1826     & 10 & $-20.17$ & U   &     	  	ESO 120-012 & 10 & $-16.72$ & A        \\
IC 1870     &  9 & $-18.27$ & A,U  &    	  	ESO 120-021 & 10 & $-15.16$ & A        \\
IC 1892     &  8 & $-18.82$ & A?,U &    	  	ESO 145-025 &  9 & $-17.05$ & A        \\
IC 1952     &  6 & $-19.37$ & U    &    	  	ESO 146-014 &  8 & $-16.44$ & A        \\
IC 1962     &  9 & $-16.65$ & TT   &    	  	ESO 149-001 &  7 & $-18.27$ & A        \\
IC 1993     &  2 & $-18.79$ & U     &   	  	ESO 149-003 & 10 & $-13.27$ & A       \\
IC 2032     & 10 & $-15.50$ & A     &   	  	ESO 150-005 &  7 & $-16.82$ & U        \\
IC 2389     &  9 & $-18.82$ & A?    &   	  	ESO 154-023 &  8& $-16.17$ & A  	 \\  
IC 2461     &$-2$ & $-21.21$ & E    &   	  	ESO 159-025 & 10& $-15.17$ & A         \\
IC 2574   &  10 & $-16.99$ & A?,E? &    	  	ESO 187-035 &  9& $-16.67$ & A  	 \\
IC 2627   &   4 & $-19.81$ & A,E  &     	  	ESO 187-051 &  9& $-16.10$ & A  	 \\
IC 2763   &   8 & $-17.01$ & C     &    	  	ESO 202-035 &  5& $-19.28$ & A  	 \\
IC 2963   &   8 & $-18.19$ & A     &    	  	ESO 202-041 &  9& $-15.49$ & A  	 \\
IC 2996   &   6 & $-18.87$ & U     &    	  	ESO 234-043 &  8& $-17.66$ & A  	 \\
IC 3102   &   0 & $-20.57$ & A,S  &     	  	ESO 236-039 & 10& $-15.84$ & A  	 \\
IC 3105   &  10 & $-15.49$ & A     &    	  	ESO 240-004 &  9& $-15.86$ & A  	 \\
IC 3155   &$-2$ & $-19.14$ & C     &    	  	ESO 245-005 & 10& $-14.38$ & U  	 \\
IC 3258   &  10 & $-16.99$ & U     &    	  	ESO 249-035 &  9& $-14.79$ & C    \\
IC 3268   &  10 & $-18.14$ & U     &    	  	ESO 249-036 &  10& $-14.62$ & U  \\
IC 3322A  &   7 & $-20.00$ & W?    &    	  	ESO 285-048 &  7& $-19.33$ & A   \\
IC 3355   &  10 & $ -9.68$ & A     &    	  	ESO 287-037 &  8& $-19.07$ & A     \\
IC 3356   &  10 & $-15.55$ & U     &    	  	ESO 289-026 &  7& $-17.70$ & A     \\
IC 3371   &   8 & $-17.23$ & E?   &     	  	ESO 289-048 &  7& $-18.12$ & E     \\
IC 3475   &  11 & $-19.18$ & U     &    	  	ESO 292-014 &  7& $-19.28$ & W?    \\
IC 3522   &  10 & $-15.46$ & A?   &     	  	ESO 293-034 & 99& $-19.70$ & A,W?,C	   \\
IC 3576   &   9 & $-16.89$ & U     &    	  	ESO 302-014 &  8& $-13.90$ & U  	 \\
IC 3583   &  10 & $-17.22$ & A,E? &     	  	ESO 302-021 &  9   & $-13.28$ & TT?	  \\
IC 3611   &   9 & $-19.02$ & A     &    	  	ESO 305-009 &  8   & $-16.79$ & U	\\
IC 3687   &  10 & $-13.75$ & A     &    	  	ESO 305-017 & 10   & $-15.66$ & TT?	\\
IC 3742   &   8 & $-17.85$ & A     &    	  	ESO 340-017 &  7   & $-19.10$ & A	\\
IC 3881   &   8 & $-17.10$ & E     &    	  	ESO 340-042 &  8   & $-18.37$ & A?,U	\\
IC 4351   &   3 & $-21.87$ & W     &    	  	ESO 341-032 &  9   & $-19.14$ & A     \\ 
IC 4214   &   0 & $-21.44$ & E     &              	ESO 342-050 &  4   & $-20.42$ & A     \\ 
IC 4407   &   9 & $-18.29$ & A     &    	  	ESO 345-046 &  7   & $-18.66$ & E?,C \\ 
IC 4468   &   4 & $-20.25$ & A     &    	  	ESO 346-014 &  7   & $-18.36$ & U     \\ 
IC 4582   &	4 & $-20.33$ & A   &	                ESO 347-008 &  9   & $ -9.69$ & U     \\
\hline
\end{tabular} 
\end{table*}

\setcounter{table}{0}
\begin{table*}
\caption{Continued.}
\begin{tabular}{lccrlccr}
\hline
Name & T-Type & Abs.3.6 $\mu$m & Classification & Name & T-Type & Abs. 3.6 $\mu$m & Classification \\
 &  & AB Mag. &  &  &  & AB Mag. &  \\
\hline
ESO 347-029 &  5   & $-18.01$ & U     &     	           ESO 505-003 &  9   & $-18.55$ &  W	     \\ 
ESO 357-007 &  9 & $-15.98$ & A       &     	   	   ESO 505-008 &  7   & $-17.65$ & E?	     \\ 
ESO 357-012 &  8 & $-17.89$ & A?,U    &     	   	   ESO 505-009 &  6   & $-16.90$ & U	     \\ 
ESO 358-015 &  9 & $-15.88$ & A?      &     	   	   ESO 505-013 &  7   & $-18.76$ & A	     \\ 
ESO 358-020 & 10 & $-17.39$ & E?      &     	   	   ESO 505-023 &  9   & $-16.06$ & A	     \\ 
ESO 358-054 &  9 & $-16.83$ & A       &     	   	   ESO 506-029 &  8   & $-18.69$ & A	     \\ 
ESO 358-060 & 10 & $-12.99$ & U       &     	   	   ESO 508-007 &  8   & $-17.42$ & U	     \\ 
ESO 358-063 &  5 & $-20.16$ & A       &     	   	   ESO 508-015 & 10   & $-18.19$ & A	     \\ 
ESO 359-003 &  9 & $-16.58$ & U       &     	   	   ESO 508-019 &  9   & $-19.04$ & A,TT?   \\ 
ESO 359-031 & 10 & $-16.45$ & A       &     	   	   ESO 508-024 &  5   & $-20.19$ & A	     \\ 
ESO 361-009 & 10 & $-15.08$ & U       &     	   	   ESO 508-030 & 10   & $-16.09$ & A,C?    \\ 
ESO 361-019 & 10 & $-18.16$ & A       &     	   	   ESO 510-026 & 10   & $-17.85$ & U	     \\ 
ESO 362-011 &  4 & $-19.86$ & A       &     	   	   ESO 510-058 &  9   & $-19.60$ & C	     \\ 
ESO 362-019 &  8 & $-16.44$ & E       &     	   	   ESO 510-059 &  5   & $-19.24$ & A,C       \\ 
ESO 399-025 &$-5$ & $-20.19$ & U      &     	   	   ESO 532-014 &  7   & $-16.13$ & A,TT    \\ 
ESO 402-025 &  9 & $-17.50$ & C       &     	   	   ESO 532-022 &  7   & $-17.93$ & A,TT    \\ 
ESO 402-026 &  2 & $-21.76$ & C       &     	   	   ESO 539-007 &  9   & $-16.79$ & U	     \\ 
ESO 404-003 &  6 & $-19.09$ & A       &     	   	   ESO 540-031 & 10   & $-11.50$ & A	     \\ 
ESO 404-017 &  9 & $-17.99$ & A       &     	   	   ESO 541-005 &  9   & $-16.58$ & A	     \\ 
ESO 404-027 &  2 & $-20.00$ & A       &     	   	   ESO 545-005 &  3   & $-19.52$ & A,E       \\ 
ESO 406-042 & 10   & $-16.72$ & A     &     	   	   ESO 545-016 &  8   & $-16.69$ & A	     \\ 
ESO 407-007 &  0   & $-20.26$ & U     &     	   	   ESO 547-012 &  9   & $-15.99$ & E?	     \\ 
ESO 407-018 & 10   & $-11.96$ & U     &     	   	   ESO 547-020 & 10   & $-16.77$ & A	     \\ 
ESO 408-012 &  9   & $-18.68$ & U     &     	   	   ESO 548-005 &  9   & $-17.57$ & U	     \\ 
ESO 409-015 & 10   & $-13.69$ & A     &     	   	   ESO 548-009 &  8   & $-17.71$ & U	     \\ 
ESO 410-012 &  9   & $-14.72$ & A     &     	   	   ESO 548-016 &  1   & $-17.35$ & TT,C   \\ 
ESO 411-013 &  9   & $-15.70$ & U     &     	   	   ESO 548-032 &  8   & $-17.46$ & A	  \\ 
ESO 420-006 & 10   & $-15.21$ & U     &     	   	   ESO 548-082 & 10   & $-16.17$ & U	  \\ 
ESO 422-033 & 10   & $-16.31$ & U     &     	   	   ESO 549-002 & 10   & $-15.82$ & U	  \\
ESO 438-017 &  7   & $-17.24$ & U     &     	   	   ESO 549-018 &  3  &  $-19.87$ & A	     \\
ESO 440-004 &  8   & $-19.09$ & A,TT  &     	   	   ESO 549-035 &  9  &  $-16.31$ & A	     \\
ESO 440-044 & 10   & $-16.37$ & E     &     	   	   ESO 550-005 &  7  &  $-16.81$ & W?	     \\
ESO 440-046 &  8   & $-17.42$ & E     &     	   	   ESO 551-016 &  8  &  $-16.56$ & U	     \\
ESO 440-049 &  5   & $-18.16$ & A     &     	   	   ESO 553-017 &  8  &  $-16.98$ & U	     \\
ESO 441-014 &  9   & $-17.81$ & C     &     	   	   ESO 567-048 &  9  &  $-14.13$ & U	     \\
ESO 443-069 &  8   & $-20.01$ & A     &     	   	   ESO 569-014 &  7  &  $-19.13$ & E	     \\
ESO 443-079 &  9   & $-16.58$ & A,E?  &     	   	   ESO 572-030 &  9  &  $-17.56$ & A	     \\
ESO 443-080 &  9   & $-18.31$ & A     &     	   	   ESO 573-003 & 10  &  $-999  $ & U	     \\
ESO 443-085 &  7   & $-17.61$ & E?    &     	   	   ESO 576-001 &  1  &  $-21.26$ & C	   \\ 
ESO 444-033 &  9   & $-17.75$ & E     &     	   	   ESO 576-005 &  7  &  $-18.47$ & U	     \\
ESO 444-037 & 10   & $-17.56$ & A     &     	   	   ESO 576-008 &  1  &  $-19.25$ & U	   \\ 
ESO 444-078 & 10   & $-999  $ & A,U   &     	   	   ESO 576-040 &  8  &  $-17.90$ & E	     \\
ESO 445-089 &  7   & $-19.96$ & A     &     	   	   ESO 576-050 &  7  &  $-19.22$ & E	     \\
ESO 462-031 &$-1$   & $-19.18$ & C    &     	   	   ESO 576-059 &  9  &  $-17.00$ & U	     \\
ESO 466-036 &$-5$   & $-20.03$ & A?   &     	   	   ESO 577-038 &  9  &  $-15.96$ & E	   \\ 
ESO 467-051 &  9   & $-16.57$ & C     &     	   	   ESO 580-034 &  7  &  $-17.59$ & A,C     \\ 
ESO 469-008 &  9   & $-16.12$ & A     &     	   	   ESO 582-004 &  2  &  $-18.07$ & U	     \\
ESO 476-010 & 10   & $-16.23$ & A     &     	   	   ESO 601-025 &  9  &  $-17.58$ & A	     \\
ESO 479-025 &  9   & $-17.83$ & I     &     	   	   ESO 601-031 &  9  &  $-17.20$ & A?,U      \\
ESO 480-020 &  5   & $-16.05$ & C,U   &     	   	   ESO 602-003 & 10  &  $-17.67$ & A	     \\
ESO 481-014 &  8   & $-17.29$ & TT?   &     	   	   PGC 143     & 10  &  $-999  $ & U	     \\
ESO 482-005 &  7   & $-17.18$ & E     &     	   	   PGC 2689    & 10  &  $-16.83$ & U	     \\
ESO 485-021 &  6   & $-17.43$ & TT    &     	   	   PGC 2805    &  9  &  $-16.45$ & A,E?      \\
ESO 486-003 & 10   & $-16.18$ & A?    &     	   	   PGC 3855    &  9  &  $-17.16$ & U	    \\
ESO 486-021 & 10  & $-15.22$ & TT     &     	   	   PGC 4143    & 11  &  $-16.99$ & U	    \\
ESO 501-079 & 10   & $ -999 $ & U     &     	   	   PGC 6244    &  9  &  $-16.48$ & A	   \\ 
ESO 501-080 &  9   & $-16.95$ & A     &     	   	   PGC 6626    &  6  &  $-17.66$ & U	  \\ 
ESO 502-016 &  9   & $-16.91$ & A     &     	   	   PGC 7900    &  8  &  $-17.64$ & A,U     \\  
ESO 502-020 &  7  &  $-17.36$ & A     &     	   	   PGC 8295    &  8  &  $-17.47$ & E	    \\
ESO 502-023 & 10  &  $-15.32$ & U     &     	   	   PGC 8962    &  10  & $-16.66$ & W?	    \\  
ESO 504-028 &  6  &  $-18.48$ & A     &     	   	   PGC 12068   &   9  & $-17.92$ & A	\\
ESO 505-002 & 10   & $-17.44$ & A     &                    PGC 12608   &	7  & $-16.71$ & A     \\
\hline
\end{tabular} 
\end{table*}
\setcounter{table}{0}
\begin{table*}
\caption{Continued.}
\begin{tabular}{lccrlccr}
\hline
Name & T-Type & Abs.3.6 $\mu$m & Classification & Name & T-Type & Abs. 3.6 $\mu$m & Classification \\
 &  & AB Mag. &  &  &  & AB Mag. &  \\
\hline
PGC 12664   &	7  & $-18.91$ & A	 & 	      PGC 45195   &  9  &  $-18.78$ & A   \\	       
PGC 12981   &	9  & $-17.88$ & A	 & 	      PGC 45359   & 10  &  $-16.22$ & A   \\	 
PGC 14487   &	9  & $-17.51$ & A	 & 	      PGC 45652   & 99  &  $-19.66$ & U   \\
PGC 14768   &	7  & $-17.63$ & E	 & 	      PGC 45958   &  5  &  $  -999$ & U   \\ 
PGC 15214   &  10  & $-16.34$ & A	 & 	      PGC 46261   &  6  &  $-20.04$ & C   \\ 
PGC 15625   &  10  & $-17.12$ & A	 & 	      PGC 47721   &  2  &  $-20.48$ & C   \\ 
PGC 16389   &  10  & $-12.90$ & U	 & 	      PGC 48087   &  5  &  $-19.12$ & TT?   \\
PGC 16784   &	8  & $-18.31$ & A,TT	 & 	      PGC 49521   &  8  &  $-17.11$ & E    \\	 
PGC 24469   &   9  & $-18.09$ & A,TT,C   & 	      PGC 51291   & 10  &  $-18.11$ & E    \\	 
PGC 28308   &$-1$  & $-21.72$ & W?	 & 	      PGC 52336   & 99  &  $-999  $ & U \\	 
PGC 29086   &	8  & $-15.14$ & E	 & 	      PGC 52809   &  7  &  $-19.95$ & A \\	 
PGC 29653   &  10  & $-13.73$ & U	 & 	      PGC 52935   & 10  &  $-17.83$ & I,C   \\
PGC 31979   &	7  & $-18.38$ & E	 & 	      PGC 52940   &  9  &  $-18.62$ & I,C   \\   
PGC 35271   &  10  & $-14.53$ & A	 & 	      PGC 53134   &  8  &  $-18.82$ & TT,I,C \\ 
PGC 36217   &	6  & $-18.94$ & A	 & 	      PGC 53634   &  8  &  $-17.97$ & E \\	 
PGC 36643   &	7  & $-18.60$ & E	 & 	      PGC 54817   & 99  &  $-18.87$ & TT  \\
PGC 37373   &	7  & $-18.26$ & U	 & 	      PGC 65367   & 10  &  $-11.25$ & U  \\    
PGC 38250   &	8  & $-17.94$ & U	 & 	      PGC 66559   &  8  &  $-18.44$ & TT,M,C  \\
PGC 41725   &	9  & $-16.80$ & E	 & 	      PGC 67871   &  7  &  $-18.21$ & U  \\  
PGC 41965   &	8   & $-18.13$ & A	 & 	      PGC 68061   &  4  &  $-19.40$ & I  \\  
PGC 42868   &	6   & $-19.30$ & A	 & 	      PGC 68771   &  8  &  $-18.99$ & E   \\ 
PGC 43341   &	9   & $-17.10$ & A	 & 	      PGC 69404   & 99  &  $-20.94$ & U  \\  
PGC 43458   &	7   & $-19.35$ & A	 & 	      PGC 69415   & 10  &  $-14.35$ & U  \\  
PGC 43679   &	7   & $-19.39$ & C	 & 	      PGC 72006   &  8  &  $-17.09$ & E  \\
PGC 43851   &  10   & $-16.41$ & U	 & 	      PGC 91228   & 10  &  $-14.25$ & U  \\   
PGC 44278   &  11  & $-17.27$ & U        & 	      PGC 91408   &  8  &  $-16.92$ & C  \\  
PGC 44532   &  10  & $-16.96$ & U        & 	      PGC 91413   &  9  &  $-17.36$ & A \\ 
PGC 44906   &  9  &  $-17.96$ & A        &	                  &     &           &   \\		    		\hline
\multicolumn{8}{l}{
  \begin{minipage}{15cm}
    \small Notes: Classification symbols are A = asymmetry, E = extension, W = warp,
TT = tidal tail, S = shell, I = interaction, M = merger, PR = polar ring, C =
companion, U = uncertain. `U' in the Classification column means that any feature 
detection was very uncertain, often due to a bright star right next to a galaxy or
a low surface  brightness light distribution that gives the galaxy a very patchy
appearance. A question mark (?) after a Classification letter means that the
classification of that feature as such was uncertain (but not due to an overall
uncertainty factor which is marked by the `U' letter). The morphological
`T'-type is based on classification in the 3.6 $\mu$m IRAC images \citep{buta14} 
and the 3.6 $\mu$m absolute AB magnitudes are from \citet{munoz14}. `99/$-999$' 
is used in the second and third columns, respectively, if the value could not 
be derived from the data, for example if the galaxies are right next to a very bright 
star, are resolved out (did not have a continous disc) or have an extremely low surface brightness.
  \end{minipage}
}\\
\end{tabular}      			    

\end{table*}

\begin{table*}

\caption{MAIN OUTER REGION STATISTICS OF S$^{4}$G GALAXIES.\label{table2}}
\begin{tabular}{lccccc}
\hline
Asym. & Unq. & Asym. & Asym. & Ext. & Warps \\
 & Asym. & Int. & Comp. &  & \\
\hline
22\%$\pm$1\% & 20\%$\pm$1\% & 29\%$\pm$7\% & 30\%$\pm$3\% & 6\%$\pm$1\% & $7\%\pm$1\% ($12\%\pm$4\%)\\
\hline
\multicolumn{6}{l}{
  \begin{minipage}{11cm}
    \small Notes: Unq. = Unquestionable; Asym. = Asymmetries; Int. = Interacting; Comp. =
Companions; Ext. = Extensions. Warp fractions are for galaxies with 
incl. $>$~$65{\degr}$ ($>$~$80{\degr}$). The third and fourth columns give the fraction
of asymmetric galaxies among interacting galaxies and among galaxies with companions,
respectively.
  \end{minipage}
}\\
\end{tabular}
\end{table*}

\subsection{Shell Galaxies}
\label{shellstats}

All but one of the eight shell galaxies have T-types from 0 to 4, i.e., S0/a to Sbc, 
thus somewhat surprisingly containing only one elliptical galaxy. This is likely to
be due to the small number of elliptical galaxies in the S$^{4}$G sample (only
46 galaxies of T-type $-4$ or less in the sample due to the selection of S$^{4}$G
galaxies by requiring a radio, most often H{\sc I}, line heliocentric radial velocity). 
On the other hand, the possible existence of shell-like features in galaxies as 
late-type as Sbc is interesting. The Sbc galaxy that has shells in the sample, 
NGC~3310, is a well-known minor merger galaxy and its shells are known 
\citep{wehner06}. The shells of NGC~474, NGC~2782, NGC~3619, and NGC~5218 are also
well known. To our knowledge, shells have not been seen in the two questionable
cases, IC~3102 and NGC~7727. In the case of NGC~2782, the shells are assumed to
be associated with a recent minor merger \citep[e.g.,][]{jogee98}. NGC~2681 is
a multiple ring galaxy \citep{buta94} and the rings may have been mistaken as 
shells. The shell galaxies in the sample have $M_{3.6}$ values of $-20.6$ -- $-21.9$, 
which correspond to less luminous than $L^{*}$ galaxies. Comparing to the
results of \citet{kim12}, we note that NGC~2634 is not officially part of the
S$^{4}$G galaxy sample, and therefore was not examined by us. In NGC~3032 the shell
structure is very faint and not visible without doing a deeper analysis
involving subtraction of the smooth light, which we did not do (most features 
of the outer region are visible in the non-smoothed versions of the image, as 
we paid attention mostly to areas outside the easily visible galaxy discs or 
spheroids). Similarly, images of NGC~5018 require unsharp-masking 
for the shell features to become clearly visible within the luminous body of
the galaxy.

\begin{figure*}
\centering
\includegraphics[width=12cm]{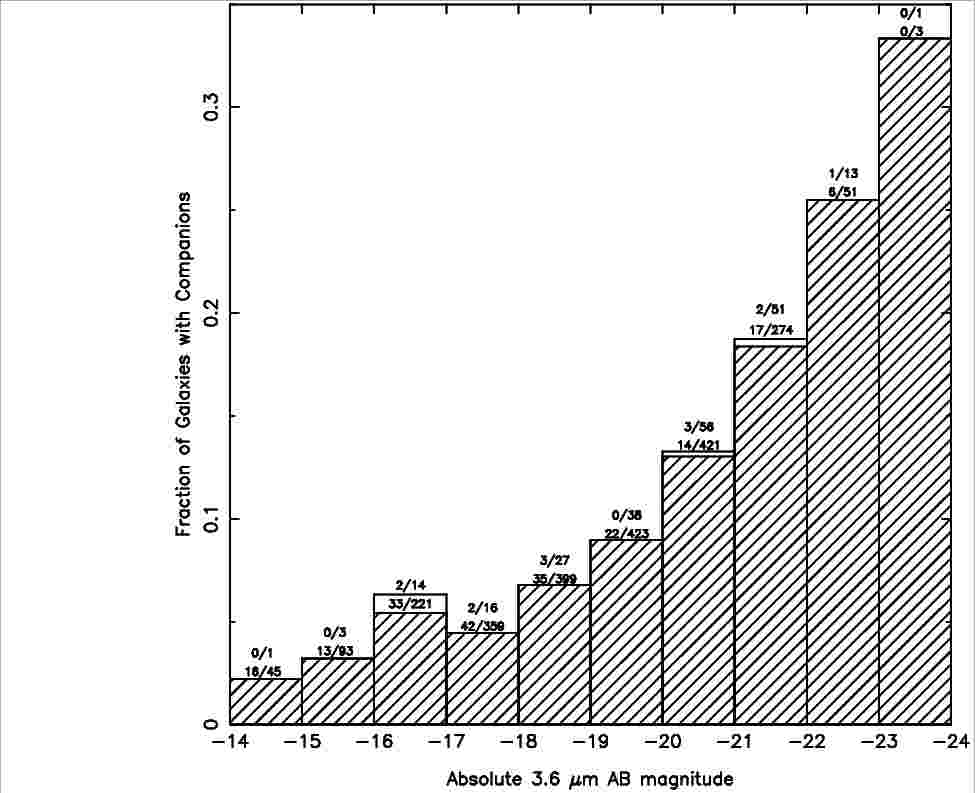}
\caption{Fraction of galaxies with companions as a function of 3.6 $\mu$m absolute AB magnitude. 
The fraction of uncertain (marked by the symbol `?' in Table~\ref{table1}) companion detections is plotted without hatching.
The fractions of galaxies with companions in a given magnitude bin with the overall uncertainty flag 'U' in
Table~\ref{table1} over all the galaxies with companions in the given magnitude bin 
(see Table~1) and of all galaxies in a given magnitude bin with the uncertainty 
flag `U' are given above the corresponding magnitude column. The luminosity could not be
determined for two galaxies with companions and 21 galaxies in the whole sample.\label{fig13}} 
\end{figure*}

\begin{figure*}
\centering
\includegraphics[width=12cm]{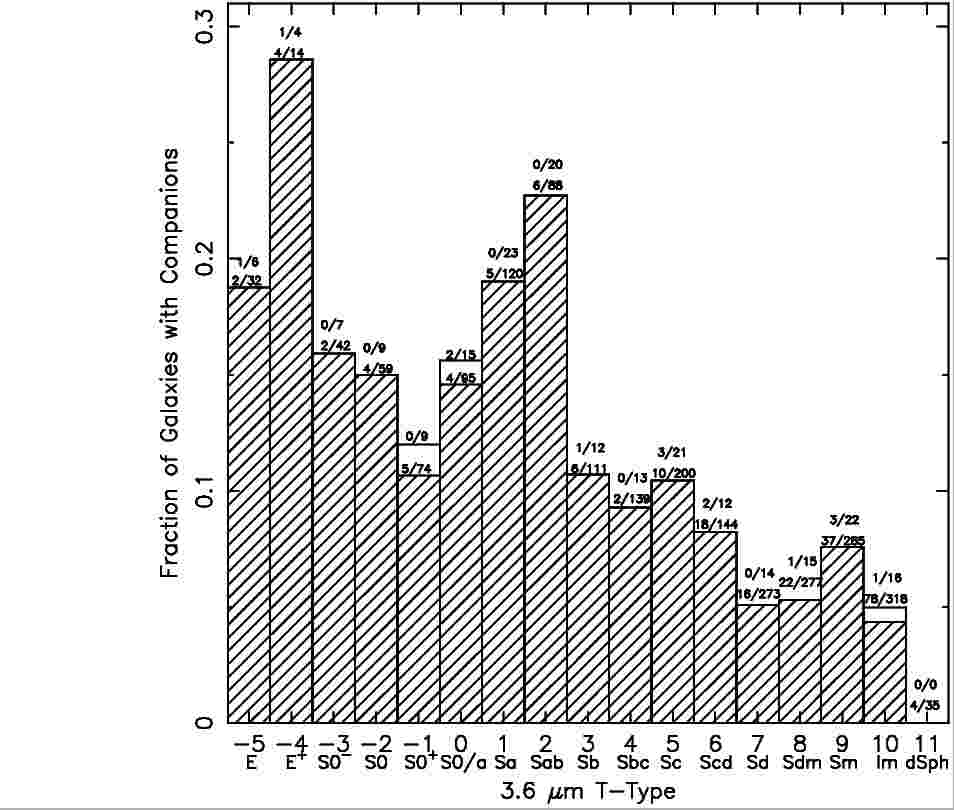}
\caption{Fraction of galaxies with companions as a function of 3.6 $\mu$m T-type. The fraction 
of uncertain (marked by the symbol `?' in Table~\ref{table1}) companion detections is plotted without hatching. The fraction of 
galaxies with companions in a given T-type bin with the overall uncertainty flag 'U' in
Table~\ref{table1} over all the galaxies with companions in the given T-type bin (see Table~1) 
and the fraction of all galaxies in a given T-type bin with the uncertainty flag `U' are 
given above the corresponding T-type column. The T-type could not be determined for four 
galaxies with companions and 18 galaxies in the whole sample. The ``dSph'' type
includes dE, dS and Sph types.\label{fig14}} 
\end{figure*}

\subsection{Polar Ring Galaxies}
\label{polarstats}

Only three polar ring candidate galaxies were detected among the S$^{4}$G 
sample galaxies. Of these, NGC~660 \citep[e.g.,][]{driel95} and NGC~5122 
\citep*[e.g.,][]{reshetnikov01} are known to be polar ring
galaxies, while NGC~681 is not known to be a polar ring galaxy. The detection
of a polar ring in this galaxy is uncertain due to the thin and large disc 
that dissects the luminous halo of this galaxy. Even an image where the underlying
disc/bulge component has been subtracted, cannot reveal with certainty whether
this feature is a ring in the galaxy plane or a polar ring. On the other hand, it is 
interesting that the well-known polar ring galaxy NGC~2685 was not detected 
in the S$^{4}$G 3.6 $\mu$m image. This is likely so because the polar ring is
actually within the main body of the galaxy when looking at it with the
histogram equalization scale in {\sc DS9}. It only shows up as a loop-like feature 
with the logarithmic intensity scale. Polar rings are also less obvious in
the older stellar population revealed at 3.6~$\mu$m. There are probably other polar rings in 
the sample at unfavorable orientations and thus they were not detected. The 
loopy features in NGC~5134 are probably part of an outer ring \citep{buta91,buta92}. 
Also, based on the findings of \citet{schweizer83} we would have expected
to find more than one polar ring among the S0 galaxies as \citet{schweizer83}
found a few per cent of all field S0 galaxies to have polar rings.

Other polar rings have been found in the S$^{4}$G sample galaxies NGC~2748, 
NGC~6870, and NGC~7465, by \citet{buta14}. Their tentative 
contours are marked in the appendices of \citet{comeron14}. However, those 
three polar rings are exceedingly subtle and/or hard to interpret. Therefore, 
they may actually not be polar rings. This is the position adopted in this paper.

\subsection{Companion Galaxies}
\label{companions}

In addition to the outer features, we also checked for nearby companions that
were visible in the 3.6~$\mu$m images (usually within about 10 arcminutes of the sample 
galaxies) by checking for their systemic velocities in NED. A galaxy was
called a companion if it was within $\pm$600~km~s$^{-1}$ of the sample galaxy
in systemic velocity. Because the images cover an arbitrary extent of space, and
more in some directions than in others, some companion galaxies were missed in
these images. We also present statistics of detected and confirmed companion
galaxies versus the 3.6 $\mu$m luminosity and the T-type in Figures~\ref{fig13}
and \ref{fig14},  respectively. Companions appear to be found most frequently
around galaxies with absolute 3.6 $\mu$m AB magnitudes around $-22$ -- $-24$,
possibly because they are the brightest galaxies in the sample and thus are
expected to have the brightest companions that are easy to detect, and 
T-types around $-4$ and 2, corresponding to Hubble classes E$^{+}$ and Sab. 

Companions have also been searched for in visible light images of S$^{4}$G galaxies 
\citep{knapen14}. However, the companion definition criteria were different. For example, the
velocity difference between the companion and the host galaxy was constrained 
to be less than $\pm$200 km~s$^{-1}$ in the 
visible light based search. In addition, the search area was more limited in the 
3.6 $\mu$m images, the 3.6 $\mu$m companion selection includes uncertain cases, 
and the visible light sample includes 477 more galaxies, so the two samples are 
not comparable (only 64 per cent of the S$^{4}$G galaxies that we classified as 
potentially having companions in the 3.6 $\mu$m images are listed 
as having companions in the visible light images). It is difficult to draw any definite 
conclusions regarding the existence of companions in the S$^{4}$G images, partly 
because of the limitations of the depth, coverage, and size of the sample.

\subsection{Comparison to Faint Features seen at Other Wavelengths}
\label{compstats}

We estimate that in our survey we pick up in several cases faint features not readily 
seen in standard shallow visible light images. Targeted deep optical imaging may go deeper 
in several other cases. For example, we do not detect the faint loops around
NGC~4013 \citep{delgado09} and NGC~5907 \citep{delgado08}, partly because
the field of view (FOV) of our IRAC observations is not large enough. 
For other galaxies, such as the eight galaxies with very faint optical features discussed
in \citet{delgado10}, who used an uncalibrated luminance filter covering most of the
visible light wavelength regime, the score is mixed. For some of the galaxies again the FOV
is relatively small (e.g., NGC~3521 and NGC~5055), while for others a few red
faint features are detected (e.g., the companion of NGC~7531 already discussed in
\citeauthor{buta87} \citeyear{buta87}, and the extensions of NGC~4651), while for
yet others we miss some of the outer features readily seen in GALEX data, which
must hence be composed of young stellar populations without counterparts in
the near-infrared (e.g., the outer disc of NGC~7531 itself).

\section{Simulation Comparison}
\label{simulations}

We have made the first attempt to utilize the information from the detected 
faint outer region features to constrain the evolution of galaxies over their 
lifetimes. We do this by comparing the outer region features in the 3.6 $\mu$m 
images to similar features around galaxies in zoom re-simulations of
cosmological galaxy evolution simulations. We have analysed a sample of 33 
simulated galaxies from \citet{martig12}. In that work each galaxy was simulated 
with a zoom re-simulation technique described in detail in \citet{martig09}. 
Star formation followed a Schmidt law with an exponent of 1.5 (above a gas density threshold 
of 0.03 M$_{\odot}$ pc$^{-3}$). \citet{martig12} also took into account kinetic supernova 
feedback and mass loss from evolved stars. The exact star formation and
feedback prescriptions probably affect the location and magnitude of 
outer disc asymmetries. We are still missing substantial physics and 
numerical resolution to fully model realistic galaxies. The spatial physical 
resolution was 150 pc, and the mass resolution 1.5 $\times 10^{4} M_{\odot}$ 
for gas and star particles, and 3 $\times 10^{5}$ for dark matter particles.

The 33 simulated haloes from \citet{martig12} are a large set of high resolution 
zoom re-simulations, with various galaxy formation histories that give rise to 
various morphologies during their evolution that is followed to $z = 0$, and therefore
they form a good set for a comparison between observed features and those forming
in simulations. The simulated galaxies were selected to have a 
mass between 2.7 $\times 10^{11}$ and 2 $\times 10^{12}$ M$_{\odot}$ at $z=0$, 
and as being in an isolated environment at $z=0$. They have a wide 
range of formation histories, from galaxies with recent major mergers 
to galaxies with no merger with a mass ratio greater than 1:10 in the 
last 9 Gyr. This resulted in a wide range of morphologies at $z=0$, even 
if 85 per cent of the sample has a bulge-to-total stellar mass ratio smaller than 0.5 
\citep[see][]{martig12}. Their final stellar mass ranges from 1.7 $\times 10^{10}$ 
to 2 $\times 10^{11}$ M$_{\odot}$.

To compare these simulated galaxies from \citet{martig12} to S$^{4}$G data, we computed mock 3.6~$\mu$m 
images for all 33 galaxies, at seven inclinations ranging from 0$\degr$ to 90$\degr$. The mock 3.6~$\mu$m images 
were computed using the PEGASE.2 stellar evolution code \citep{fioc99}, assuming 
a Kroupa initial mass function from 0.1 to 120 M$_{\odot}$ (we did not include 
any dust contribution because at 3.6 $\mu$m dust is not expected to play a 
significant role and can be excluded in the modelling). Each image corresponds 
to 100 x 100 kpc in size (with a similar total depth of 100 kpc). Each pixel 
represents 143 x 143 pc, which in terms of IRAC pixels corresponds to a galaxy 
situated 24 Mpc away (the S$^{4}$G sample galaxies are at distances of 1 to 60 Mpc,
although almost all of them are within 40 Mpc). 
We added an expected background of 0.15 MJy/sr which corresponds to a medium 
background level in IRAC images. We then converted the image into mock IRAC 30 
second frames in electron units. Next we generated a Poisson variate for the flux 
value from {\sc IDL}'s {\sc randomu} function, and added in the readnoise contribution 
which is 14.6 electrons. We then added noise to an artificial skydark, using a 
typical median value of 0.05 MJy/sr or 38 electrons, and subtracted it from the 
noise-added galaxy image. Finally we converted the image to MJy/sr, and made eight 
realizations of these images, corresponding to the IRAC observing depth, and took 
the median of them to form the final image. We selected each of the 33 simulated
galaxies in a random order but with such inclinations that the observed S$^{4}$G
galaxy inclination distribution was reproduced. We then classified the inclined
galaxy mock images in exactly the same way as we classified the S$^{4}$G IRAC 
channel 1 images. 

\begin{figure}
\centering
\includegraphics[width=8cm]{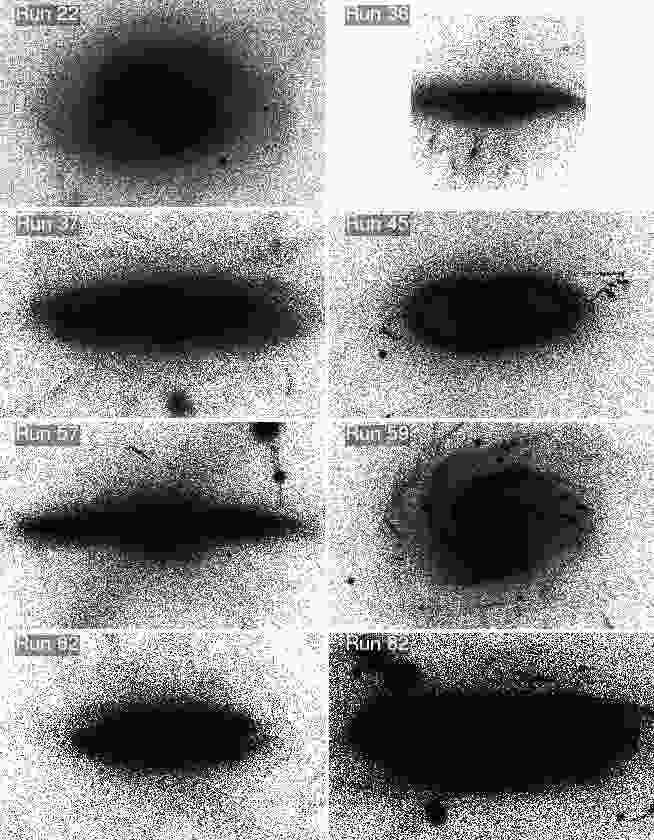}
\caption{Images of outer disc asymmetries in simulated galaxies. 
Images of all the detected asymmetries are available in the online version of 
the Journal.\label{fig15}} 
\end{figure}

\begin{figure}
\centering
\includegraphics[width=8cm]{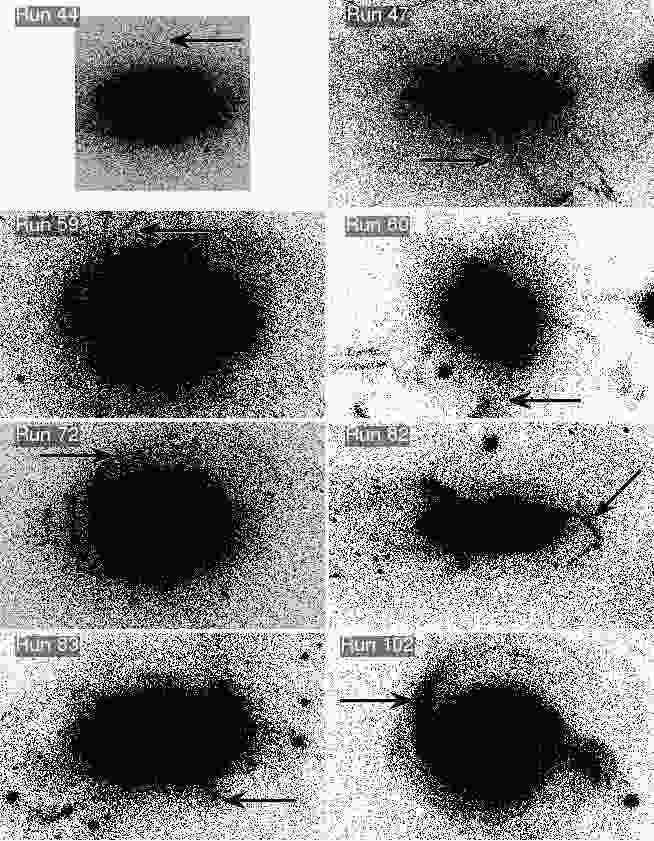}
\caption{Images of outer disc extensions in simulated galaxies.
Images of all the detected extensions are available in the online version of 
the Journal.\label{fig16}} 
\end{figure}

The detected outer disc features in the simulated galaxies are shown in Figures~\ref{fig15}
and \ref{fig16}. We have looked at the time series of simulations. It is often
not obvious what the causes of the outer disc asymmetries or extensions are,
but probable culprits include ongoing interactions, asymmetric spiral structure 
resulting from a simple companion galaxy fly-by 2--4 Gyrs ago (comparable to the age
estimate of visible interaction signs from abnormal colours or fine structure by 
\citeauthor{schweizer92} \citeyear{schweizer92}; however, it should be noted
that such a flyby does not always result in detected asymmetric structure),
`chaotic' disc reformation after a recent merger (the old disc was destroyed
by the merger, and a new disc is still in the process of settling
down and therefore it appears asymmetric or has extensions), and long-lived (5 Gyrs or
more) asymmetric spiral structure (this may be related to asymmetries in
external gas accretion). The best way forward is to statistically compare
large samples of observed and model galaxies to explore a range of possible
origins. Future models may also provide other clues to the origin of these
features, such as colours, clumpiness measurements, etc.

We find asymmetric outer discs in 11/33 galaxies or 33$\pm$8 per cent, and we find a similar
fraction of outer disc extensions (33 per cent). These numbers are higher than the
fractions of outer disc asymmetries and extensions in the S$^{4}$G sample.
Possible reasons for this discrepancy include the fact that the simulation
sample consists of mostly late-type disc galaxies, whereas the S$^{4}$G sample
has several earlier-type galaxies. As seen in Figure~\ref{fig10}, 
asymmetries are more prevalent (around 25 per cent) among late-type disc galaxies
than in the S$^{4}$G sample as a whole. Also, our current inability to model the
physics of star formation is likely to affect details of 
asymmetry formation. For example, if the star formation gas density threshold used in the
simulations was higher, then the visible disc would be detected to smaller radii
where the galaxy is more symmetric. Therefore, the simulated galaxies may allow 
star formation further out than real galaxies. Put another way, the fraction of
asymmetries in the outer regions of real galaxies may be 33 per cent, but without 
star formation in the outer parts, a fraction of the asymmetries would not be 
visible. Also, outer discs may have an additional condition for star formation, 
other than a critical density, such as needing to form molecules at low metallicity.
In fact, anything that makes a real galaxy less able to form stars in the far outer 
part than the simulated galaxy would seem to lower the asymmetry fraction for real 
galaxies. 

The reason for the high fraction of extensions in the simulated 
sample is less clear, but it could mean that some parameters in the simulations 
require further adjustments to make the simulated galaxies look more realistic 
as a whole. We found a companion and an interaction in only one of the 33 galaxies 
(this is partly a selection effect because the simulated galaxies were selected 
to be isolated at $z = 0$).

\section{Conclusions}
\label{conclusions}

This paper presents discoveries and classifications of near-infrared-detected
stellar features outside the main bodies of galaxies (at and outside of $R_{25}$) 
in the complete sample 
of 2,352 S$^{4}$G galaxies. The detected features include asymmetries, extensions, 
polar rings, warps, shells, tidal tails, and interaction/merger morphologies. 
We also tabulate nearby companion galaxies, confirmed by a reasonable systemic velocity
difference of $\pm$600~km~s$^{-1}$ in NED, as seen in the 3.6 $\mu$m images. This
list of outer disc features is conceived to be an important data base for future quantitative
studies of them when higher S/N observations become available.

We also give statistics on the features we detected. The fraction of asymmetric
galaxies in the S$^{4}$G sample is about 20 per cent. If the $\sim$~20 per cent fraction of galaxies 
with asymmetries in their outer discs is overwhelmingly due to interactions, it may imply that half of 
all galaxies have interactions that leave visible signs for $\sim$~4 Gyrs after the beginning 
of the interaction. However, an internal origin for some of these asymmetries is also 
possible, e.g., due to dark halo asymmetry induced lopsidedness. We found that the number 
of asymmetric galaxies increases with T-type, peaking in late Hubble types (T-types 5--10), 
as would be expected, because the later type galaxies are more susceptible to disturbances due
to their kinematics and stellar distributions. Surprisingly, we find shells in galaxies 
of fairly late T-types, although shells are commonly believed to be primarily features of 
early-type galaxies. 

In a first attempt to utilize our faint outer feature detections to constrain galaxy evolution 
on billions of years time scale,
we have also classified galaxies in cosmological zoom re-simulations as seen at $z = 0$, and
converted to IRAC-like images. We find a larger outer disc asymmetry fraction (by a factor of 1.5) 
in the simulated galaxy sample than in S$^{4}$G, which may be due to selection effects and
our incomplete understanding of star formation thresholds. The
simulations suggest interactions and mergers, asymmetric external gas accretion,
unfinished disc reformation, and asymmetric spiral structure as causes for asymmetry.
However, it is difficult to quantify the relative importance of these effects. Finally,
the simulations suggest that asymmetries may be visible for at least 4 Gyrs after an
interaction or merger.

\section*{Acknowledgements}

We thank Ramin Skibba for his helpful comments on a draft of this paper. We
also thank Carrie Bridge for discussions on the merger rate. We acknowledge
the helpful discussion with Chris Lintott about biases in classification.
We acknowledge financial support to the DAGAL network from the People Programme
(Marie Curie Actions) of the European Union's Seventh Framework Programme
FP7/2007-2013/ under REA grant agreement number PITN-GA-2011-289313. This work
was co-funded under the Marie Curie Actions of the European Commission
(FP7-COFUND). We also gratefully acknowledge support from NASA JPL/Spitzer grant
RSA 1374189 provided for the S$ ^{4} $G project. E.A. and A.B. thank the CNES
for support. K.S., J.--C. M.--M., T.K., and T.M. acknowledge support from the
National Radio Astronomy Observatory, which is a facility of the National
Science Foundation operated under cooperative agreement by Associated
Universities, Inc. The authors thank the entire S$ ^{4} $G team for their
efforts in this project. This work is based on observations made with the
Spitzer Space Telescope, which is operated by the Jet Propulsion Laboratory,
California Institute of Technology under a contract with NASA. Support for this
work was provided by NASA through an award issued by JPL/Caltech. We are grateful 
to the dedicated staff at the Spitzer Science Center for their help and support 
in planning and execution of this Exploration Science program. This research has 
made use of the NASA/IPAC Extragalactic Database (NED) which is operated by JPL, 
Caltech, under contract with NASA.

\bsp

\label{lastpage}

\end{document}